\documentclass[journal=langd5,manuscript=article,
layout=twocolumn,
a4paper,10pt]{achemso}

\usepackage[version=3]{mhchem} % Formula subscripts using \ce{}
\usepackage[T1]{fontenc}       % Use modern font encodings
\usepackage{graphicx,setspace}
\usepackage{amsfonts,amsmath,amssymb,bm,esint}
\usepackage[usenames]{color}
\usepackage{notes2bib}
\usepackage{mathtools}
\usepackage{tikz}
\usepackage{MnSymbol}
\usepackage{multirow}
\usepackage{chngcntr}
\usepackage{titlesec}
\usepackage{cprotect}

\newcommand{\tikzcircle}[2][red,fill=red]{\tikz[baseline=-0.5ex]\draw[#1,
radius=#2] (0,0) circle ;}
\newcommand{\tikzsymbol}[2][circle]{\tikz[baseline=-0.5ex]\node[inner 
sep=2pt,shape=#1,draw,#2]{};}

\author{Sayan Das}
\affiliation{Max-Planck-Institut f\"{u}r Intelligente 
Systeme, Heisenbergstr. 3, D-70569, Stuttgart, Germany}
\email{sayan.das@is.mpg.de}
\author{Zohreh Jalilvand}
\affiliation{Department of Chemical Engineering, City College of the City 
University of New York (CUNY), 140th St. \& Convent Av., New York, New 
York 10031, USA}
\email{zjalilv000@citymail.cuny.edu}
\author{Mihail N. Popescu}
\affiliation{Max-Planck-Institut f\"{u}r Intelligente Systeme, 
Heisenbergstr. 3, D-70569, Stuttgart, Germany}
\author{William E. Uspal}
\affiliation{Department of Mechanical Engineering, University of Hawai'i at 
M{\=a}noa, 2540 Dole Street, Holmes Hall 302, Honolulu, Hawai'i 96822, USA}
\author{Siegfried Dietrich}
\affiliation{Max-Planck-Institut f\"{u}r Intelligente Systeme, 
Heisenbergstr. 3, D-70569, Stuttgart, Germany}
\alsoaffiliation{IV. Institut f\"ur Theoretische Physik, 
Universit\"{a}t Stuttgart, Pfaffenwaldring 57, D-70569 Stuttgart, Germany}
\author{Ilona Kretzschmar}
\affiliation{Department of Chemical Engineering, City College of the City 
University of New York (CUNY), 140th St. \& Convent Av., New York, New 
York 10031, USA}
\title{
Floor- or ceiling-sliding for chemically active, gyrotactic, 
sedimenting Janus particles 
}

\begin{document}

% \begin{tocentry}
% \begin{center}
%\hspace*{0.1 truecm}\includegraphics[height = 3.2 truecm]{toc_v8.pdf}\hfill\\
%coexistence of sliding states for a chemically active Janus particle 
%All should fit within 9 cm x 3.5 cm...
%\end{center}
% Some journals require a graphical entry for the Table of Contents.
% This should be laid out ``print ready'' so that the sizing of the
% Inside the \texttt{tocentry} environment, the font used is Helvetica
% 8\,pt, as required by \emph{Journal of the American Chemical
% The surrounding frame is 9\,cm by 3.5\,cm, which is the maximum
% permitted for  \emph{Journal of the American Chemical Society}
% graphical table of content entries. The box will not resize if the
% content is too big: instead it will overflow the edge of the box.
% 
% This box and the associated title will always be printed on a
% separate page at the end of the document.
% 
%\end{tocentry}

\begin{abstract}

{\small Chemically active particles achieve motility without external forces 
and torques (``self-propulsion'') due to catalytic chemical reactions at  their 
surfaces, which change the chemical composition of the surrounding solution 
(called ``chemical field``) and induce hydrodynamic flow of the solution. 
By coupling the distortions of these 
fields back to its motion, a chemically active particle experiences an 
effective interaction with confining surfaces. This coupling can lead to a rich 
behavior, such as the occurrence of wall-bound steady states of 
``sliding''.

Most active particles are density mismatched with the solution, and 
thus tend to sediment. Moreover, the often employed Janus spheres, which 
consist of an inert core material decorated with a cap-like, thin layer of a 
catalyst, are gyrotactic (i.e., ``bottom-heavy''). Whether or not 
they may exhibit sliding states at horizontal walls depends on the interplay 
between the active motion and the gravity-driven sedimentation and 
alignment, such as the gyrotactic tendency to align the axis along the 
gravity direction being overcome by a competing, activity-driven alignment 
with a different orientation. It is therefore important to understand and 
quantify the influence of these gravity-induced effects on 
the behavior of model 
chemically active particles moving in the vicinity of walls.

For model gyrotactic, self-phoretic Janus particles, here we study 
theoretically the occurrence of sliding states at horizontal planar walls 
that are either below (``floor'') or above (``ceiling'') the particle. We 
construct ``state diagrams'' characterizing the occurrence of such 
states as a function of the sedimentation velocity and of the gyrotactic 
response of the particle, as well as of the phoretic mobility of the particle. 
We show that in certain cases sliding states may emerge \textit{simultaneously} 
at both the ceiling and the floor, while the larger part of the 
experimentally relevant parameter space corresponds to particles that 
would exhibit sliding states only either at the floor or at the ceiling -- or 
there are no sliding states at all. These predictions are critically compared 
with the results of previous experimental studies, as well as with our 
dedicated experiments carried out with Pt-coated, polystyrene-core or 
silica-core Janus spheres immersed in aqueous hydrogen peroxide solutions.

\vspace*{0.2in}
\noindent $^*$ \textit{Corresponding authors}
}

\end{abstract}
\newpage

\section{Introduction \label{intro}}

Since the first reports of self-motility for micrometer-sized colloids 
appeared fifteen years  ago \cite{Paxton2004,Fournier-Bidoz2005}, the 
topics of active particles, active fluids, and active matter have witnessed a 
rapid growth of scientific interest.  A wide variety of such particles, which 
are capable of moving autonomously, i.e., in the absence of external forces or 
torques acting on them or on the fluid, within a liquid environment -- by 
promoting chemical reactions involving their surrounding solution -- has 
been proposed and studied experimentally; see the thorough and insightful 
reviews provided in Refs. \citenum{Ebbens2010,Posner2017,Sanchez2015,SenRev}. 
One of the often 
encountered experimental realizations is that of spherical, axisymmetric 
Janus colloids, which self-propel when immersed in an aqueous hydrogen 
peroxide ($\mathrm{H_2 O_2}$) solution. These particles are obtained by 
depositing a thin film of catalyst material over a spherical core of a material 
without catalytic properties. \bibnote{The deposition of the main, catalytic 
material is sometimes preceded by that of an ultrathin layer of a material, 
which facilitates the adhesion of the catalyst, such as Ti on polystyrene before 
depositing Pt (see, e.g., Ref. \citenum{Baraban2012}).} Typical realizations, 
which will be of main interest for the present study, are Pt on polystyrene 
(Pt/PS) particles 
\cite{Howse2007,Baraban2012,Ebbens2013,Brown2014,Takatori2016,Ilona2018} and 
Pt on silica (Pt/$\mathrm{SiO_2}$) particles
\cite{Stocco2015,Simmchen2016,Uspal2018b,diLeonardo2016,Katuri2018}; another 
example, which is recently attracting much interest, is that of  
titania on silica ($\mathrm{TiO_2}$/$\mathrm{SiO_2}$) particles 
\cite{Uspal2018c,Mark17}, for which the decomposition of $\mathrm{H_2 O_2}$ 
occurs via photocatalysis upon illumination with UV light of a suitable 
wavelength. The reviews in Refs. 
\citenum{Sanchez2015,SenRev}, and \citenum{Palacci2017} provide detailed 
accounts and 
exhaustive lists of references of the numerous other types (with respect to 
core and catalyst) of Janus particles reported in experimental studies, as well 
as of the various mechanisms of self-motility.

Irrespective of the exact mechanism of motion, the motility of chemically 
active Janus particles is connected with  hydrodynamic flow of the solution as 
well as with a spatially varying distribution of chemical species 
(``chemical field'') around the particle. When such particles operate in the 
vicinity of 
walls, or, in general, near any type of spatially confining boundary, their 
chemical and hydrodynamic fields are distorted due to the impenetrability of 
the wall and due to the no-slip hydrodynamic condition at the wall. The 
coupling of these disturbances back to the particle influences its motion, i.e., 
the particle experiences an effective interaction with the confining boundary 
(see, e.g., the review in Ref. \citenum{Popescu2018a}). For axially symmetric 
Janus colloids, one of the most striking consequences of these interactions is 
the occurrence of steady-states of sliding along the wall, in which the 
particle moves parallel to the wall while maintaining a constant distance from 
it and a constant orientation of its axis relative to the direction normal to 
the wall. Such states, predicted theoretically in Ref. \citenum{Uspal2015a} for 
particles moving by self-phoresis, have been observed in experiments with Janus 
particles 
\cite{volpe11,Baraban2012,Simmchen2016,Brown2014,Takatori2016,Ebbens2019,
Isa2017,Uspal2018c,Mark17,Katuri2019} and provide the rationale behind concepts 
such as ``guidance'' by topographical features 
\cite{Simmchen2016,Howse2015,Katuri2018,Katuri2019} or rheotactic behavior of 
spherical Janus colloids \cite{Uspal2015b,Uspal2018a}. (Note that the 
experiments in Refs. \citenum{Takatori2016} and \citenum{Katuri2019}, and 
a part of those in Ref. \citenum{Isa2017}, involve setups with 
chemically 
active Janus particles in the vicinity of a fluid interface; in such cases, 
additional contributions to the sliding state, emerging from, e.g., induced 
Marangoni flows \cite{Alvaro2016}, are expected to play a role.)

A common characteristic of the spherical Janus colloids discussed above is 
that they are ``bottom heavy'', owing to the mismatch in density between 
the catalyst layer and the core material. This bottom heaviness leads to a 
gyrotactic response, 
i.e., a torque that rotates the axis of the particle towards alignment 
with the direction of gravity; accordingly, such a Janus particle immersed 
in solution exhibits a preferred orientation with the cap of catalyst 
(typically the denser material) pointing into the direction of gravity 
\cite{Ebbens2013}. (Note that this effect occurs in addition to the 
emergence of polar alignment along the gravity direction, as reported in 
Refs. \citenum{Stark2011} and \citenum{Ginot2018b}, which is due to the 
interplay between sedimentation, self-motility, and thermal fluctuations. 
Furthermore, an interplay between self-propulsion, sedimentation, and shape 
lacking axial symmetry can also lead to very complex patterns of gravitaxis or 
cross-gravitaxis, i.e., motion orthogonal to the direction of gravity, as 
reported by studies with L-shaped particles 
\cite{Bechinger2014}.) Such particles have the potential to mimic the behavior 
of ``bottom heavy'' motile micro-organisms, such as 
\textit{Chlamydomonas}, which are known to exhibit very interesting collective 
dynamics via the interplay of self-motility with the gyrotactic and gravitactic 
response to gravity as an external aligning field. Theoretical studies of 
heavy and bottom-heavy hydrodynamic squirmers have been successful in capturing 
phenomena exhibited by swimming microorganisms, such as the ``dancing'' of 
\textit{Volvox} algae \cite{Goldstein2009}. Collective alignment and 
self-assembly into spinners have been reported for monolayers of heavy squirmers 
\cite{Lintuvuori2019}, and the stability of such monolayers has been 
thoroughly analyzed \cite{Pedley2019}. Finally, we note that recently the 
collective sedimentation of squirmers under gravity \cite{Stark2017} and the 
sedimentation of a heavy squirmer  in the vicinity of a wall 
\cite{Stark2018} have been studied.

\begin{figure}[!htb]
    \centering
    \includegraphics[width=.99\columnwidth]{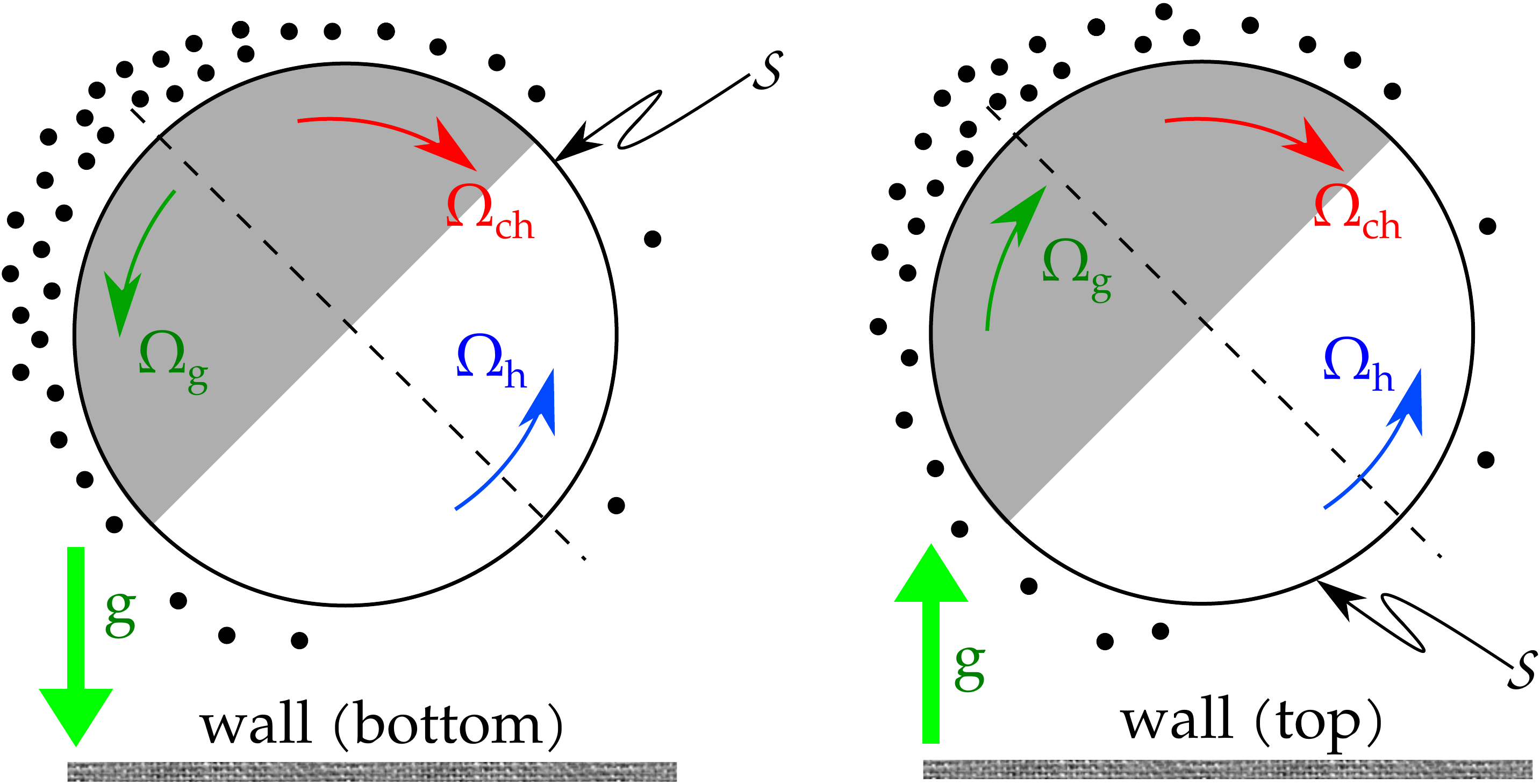}
     \caption{
\label{fig1}
{\small 
Schematic illustration of a spherical, chemically active, gyrotactic 
(bottom heavy) Janus particle approaching (left panel) a bottom wall (``floor'') 
or (right panel) a top wall (``ceiling''). As depicted in the right panel, the 
configuration of approaching a ``ceiling'' is similar to that of approaching a 
``floor'', except for a reversal of the direction of gravity (and thus of 
the gyrotactic response, i.e., the green arrows). The use of this equivalence 
is advantageous because the definition of the configuration of the particle 
(i.e., the height above the wall and the orientation with respect 
to the wall) involves in both cases the same variables $h$ and $\theta$ (see the 
main text). The chemical activity is modeled as the release, at the catalyst 
side (dark gray area) of the surface $\mathcal{S}$ of the particle, of solute 
molecules (black dots) into the surrounding solution. (As depicted in the 
figure, the distributions of solute at a given configuration ($h$, $\theta$) are 
the same for a particle below the ceiling or above the wall.) The rotations due 
to hydrodynamic (blue arrow) and chemical (red arrow) effective 
interactions with the wall (i.e, the coupling of the wall-induced distortions of 
the chemical and the flow fields to the motion of the particle), 
respectively, are shown for the case in which they oppose each other, and thus 
could eventually balance in the absence of gravity (as in the case, e.g., of a 
sliding state along a vertical wall).
}
}
\end{figure}

Returning to the case of chemically active particles and their dynamics in 
the vicinity of a wall, we note that in the absence of 
gravity-induced effects the chemical and hydrodynamic interactions of the 
active Janus particle with 
the wall lead to an orientation of the axis of the Janus particle, which is 
typically within $\pm 30^\circ$ from being parallel to the wall 
\cite{Uspal2015a,Koplik2016,Bayati2019}. In the presence of gravity, a bottom 
heavy Janus particle experiences in addition the competing effect of the 
gyrotactic preference for alignment of its axis along the  direction of gravity, 
which is perpendicular to a horizontal wall; see Fig. \ref{fig1}. (Obviously, 
any particle, which is not density matched with the liquid, experiences 
also a gravitational force.) Accordingly, 
this 
raises the question of whether, for model chemically active Janus particles 
and upon accounting for the effects due to gravity, sliding states along 
a horizontal wall still 
occur.
%%%%%%%%%%%%%%%%%%%%%%%%%%%%%%%%%%%%%%
%%%%%%%%%%%%%%%%%%%%%%%%%%%%%%%%%%%%%%

For particles that sediment to the floor, there is ample experimental 
evidence for sliding states 
\cite{Simmchen2016,Uspal2018c,Katuri2018,Baraban2012}. Theoretical 
studies, which employ simple models of self-phoretic, gyrotactic 
Janus particles, capture this phenomenology \cite{Simmchen2016,Uspal2018c}. For 
example, it has been shown that, for parameters corresponding to a typical 
Pt/$\mathrm{SiO_2}$ particle, (i) the model self-phoretic particle employed 
in Ref. \citenum{Uspal2015a} -- augmented with terms accounting for 
sedimentation and bottom heaviness -- predicts the emergence of sliding 
states; and (ii) for the sliding state, the rotation of the axis due to the 
gravitational torque, and the rotations due to the chemical and hydrodynamic 
interactions of the particle with the wall, are all comparable in magnitude.

The correlation of the latter observation with the changes of sign within the 
effects of gravity at the ceiling and at the floor (e.g., compare the gyrotactic 
response, indicated by green arrows, in the panels of Fig. \ref{fig1}), leads 
to an important conclusion: the question concerning the emergence of sliding 
states cannot be answered based solely on the behavior at the floor, even if the 
latter is completely understood. Accordingly, an additional study is required in 
order to understand if the models employed in the theoretical analysis are also 
compatible with the experimental observations of sliding states occurring at 
upper planar boundaries 
(ceiling)\cite{Brown2014,Takatori2016,Howse2015,Ebbens2019}.

Additional motivation for such a study arises from the question of how 
active particles, which are both heavy (i.e., they sediment to the floor in the 
absence of chemical ``fuel'') and bottom-heavy (i.e., there is a gravitational 
torque due to the catalytic cap), may reach a state of sliding at the ceiling. 
One scenario, applicable for active particles moving with the active cap at the 
back, is the following. At sufficiently strong chemical activity, and thus at 
high velocity of self-propulsion, the particles lift off from the floor 
and, owing to the bottom-heavy gyrotactic response aligning their caps down, 
exhibit a persistent three-dimensional motion against gravity (see, e.g., the 
experimental reports in Refs. 
\citenum{Brown2014,Ebbens2013,Uspal2018c,Ebbens2019}). This 
leads to collisions with the ceiling, which is a prerequisite for the emergence 
of a sliding state\cite{Uspal2015a,Simmchen2016,Popescu2016}; obviously, in this 
regime of chemical activity the sliding states at the floor cease to exist. The 
issue is whether the above lift-off scenario is actually a necessary 
condition 
for the occurrence of a sliding state at the ceiling. For example, one can think 
of an alternative scenario in which active particles sediment in the bulk, 
i.e., far from confining boundaries gravity dominates self-propulsion, but a 
state of sliding at the ceiling exists. In this scenario, an active 
particle, which is initially located within the basin of attraction of 
that state, will end 
up sliding along the ceiling, even though the chemical activity is not strong 
enough to ensure a lift-off from the floor. This scenario, that for given 
material properties -- e.g., Pt/PS particles of a given radius and a given 
chemical activity, i.e., a fixed concentration of H$_2$O$_2$ -- sliding states 
at the floor coexist with sliding states at the ceiling, so far has not been 
explored theoretically. Yet, it would rationalize the observation, in the 
context of a single experiment, of motile Janus particles both at the top 
and at the bottom walls. Such a situation can be inferred from the studies in 
Refs. 
\citenum{Brown2014} and \citenum{Howse2015}.

The discussion above illustrates a complex interplay between the chemical 
activity of the Janus particle, the effects due to gravity (sedimentation and 
gyrotactic alignment), and the vertical location of the bounding wall with 
respect to the particle, i.e., whether it is above or below the particle. 
This can affect both qualitatively and quantitatively the emergent 
dynamical behavior. It is therefore important to understand and quantify the 
influence of these gravity-induced effects on the behavior 
of model chemically 
active particles moving in the vicinity of walls. Here, we study both 
theoretically and experimentally, for model bottom heavy, 
self-phoretic Janus particles, the issue of the occurrence of sliding states at 
planar, horizontal floor and ceiling walls. We use a simple model of 
self-phoretic motion \cite{Golestanian2007}, which has been previously shown to 
capture qualitatively the behavior exhibited in experimental studies of Janus 
particles near walls \cite{Simmchen2016,Katuri2018,Uspal2018c}. For two choices 
of the chemical activity of the particle, which are typically invoked in 
theoretical studies 
\cite{Golestanian2007,Golestanian2014,Popescu2018b,Liverpool2017}, we 
construct ``state diagrams''. They summarize the behavior of the  model 
gyrotactic, self-phoretic Janus particles near walls as functions of the 
relevant parameters of the dynamics. These parameters are the ratios 
between activity- and gravity-induced translational and rotational velocities, 
and the ratio of the 
phoretic mobilities of the active and inert parts of the Janus particle. The 
theoretical predictions are critically compared with the results of the present 
experiments conducted with Pt/PS and Pt/SiO$_2$ Janus particles immersed in 
aqueous H$_2$O$_2$ solutions, as well as with previously published experimental 
studies in which sliding states at the ceiling have been reported 
\cite{Brown2014,Howse2015,Ebbens2019}.

\section{Experimental section \label{exper}}

Polystyrene latex particles of two sizes, radii $R = 1.2~\mathrm{\mu 
m}$ and $2.5~\mathrm{\mu m}$, respectively, as well as silica particles of $R 
= 0.5~\mathrm{\mu m}$ and $2~\mathrm{\mu m}$, respectively, are employed in the 
study. Janus particles are fabricated by using the physical vapor 
deposition (PVD) technique to coat the spherical polystyrene (or silica) 
particles with platinum. In brief, for each type and size of particle, a 
monolayer of them is assembled onto a pre-cleaned glass slide using the 
convective assembly method \cite{Velev2004}. Subsequently, these monolayers of 
particles are 
exposed to platinum vapor in a PVD machine (Ted Pella), and a quantity of 
Pt -- equivalent to a planar layer of thickness $\delta_{rep} \simeq 
9~\mathrm{nm}$ (as \textit{rep}orted by the device) -- is deposited onto 
such a monolayer of particles.  After the Pt deposition, the Janus particles 
formed this way are dispersed in deionized water by sonicating the glass slides 
for one minute.

%%%%%%%%%%%%%%%%%%%%%%%%%%%%%%%%%%%%%
\begin{figure}[!b]
\centering
\includegraphics[width=.99\columnwidth]
{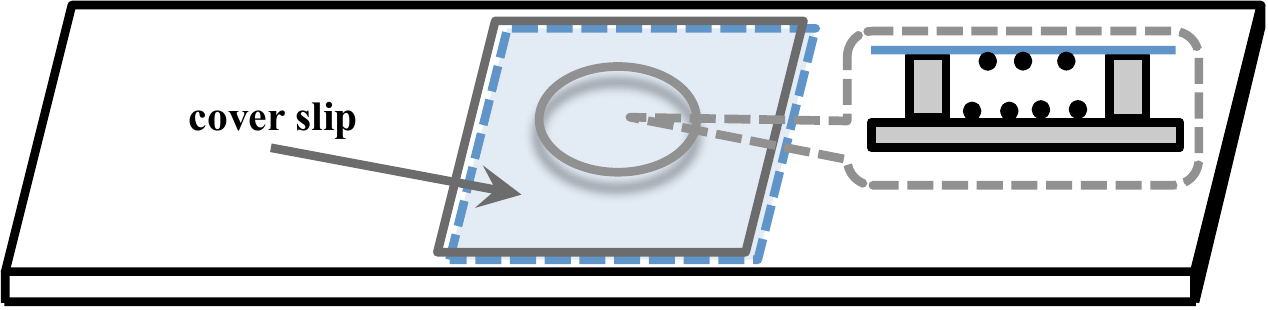}
\caption{
\label{fig2}
{\small 
Sketch of the experimental cell (height of $\approx 500~\mathrm{\mu m}$, volume 
of $\approx 40~\mathrm{\mu l}$). The inset illustrates the vertical cross 
section of the cell, with the black dots depicting active particles at the top 
(blue) and bottom (gray) glass slides.
}
}
\end{figure}
%%%%%%%%%%%%%%%%%%%%%%%%%%%%%%%%%%%%%
The experimental cell used to study the motion of the particles, while 
they are suspended in an aqueous $\mathrm{H_2O_2}$ solution, is composed of a 
glass slide and a silicone isolator ring (Invitrogen Corp.). The 
silicone ring is attached onto a pre-cleaned glass slide with the help of 
an adhesive added onto its bottom surface, creating a cylindrical well of 
approximately 9 mm diameter and 0.5 mm depth (see Fig. \ref{fig2}). The 
large height of the cell (at least 100 particle diameters, for the particles 
employed in our study) ensures that the dynamics of an active particle in the 
vicinity of one of the walls (top or bottom) is, to a very good approximation, 
unaffected by the presence of the other, distant wall.

The concentration of the aqueous dispersion of Janus particles is adjusted 
to $\lesssim 1 \%$ volume fraction; the use of a very dilute particle 
suspension ensures that the experiments are exploring the regime of ``single 
particle motion'', and it also reduces the probability of forming bubbles 
(which would give rise to spurious motions of nearby particles). Subsequently, 
the particle suspension is mixed with an adequate volume of aqueous hydrogen 
peroxide solution, in order to obtain the desired $\mathrm{H_2O_2}$ 
concentration, as follows. First, an aqueous solution of 6\% (v/v) concentration 
$\mathrm{H_2O_2}$ is prepared from an aqueous stock solution of 30\% 
$\mathrm{H_2O_2}$ (Fisher Scientific). A volume of the dispersion of Janus 
particles is then gently mixed with an equal volume of the aqueous 6\% (v/v) 
$\mathrm{H_2O_2}$ solution, resulting in a 3\% (v/v) $\mathrm{H_2O_2}$ 
suspension of Janus particles. Subsequently, the well-mixed solution is 
transferred to the well by means of a pipette, and the well is carefully covered 
with a cover slip in order to avoid disturbances due to air currents and 
to reduce water evaporation. Moreover, the top cover slip provides the 
``ceiling'' wall for studying the eventual emergence of sliding states at the 
top. 

Since the self-propelled motion of the particles starts immediately upon mixing 
of the particle suspension with the aqueous $\mathrm{H_2O_2}$ solution, the 
particles are already motile when the mixed solution is placed into the 
experimental well. Consequently, this setup provides optimal conditions for the 
eventual occurrence of sliding states at the upper cover slip, because the 
distribution of motile particles within the volume is relatively uniform at the 
beginning of the experiment.

The motion of the Janus particles is observed using an Olympus BX-51 microscope 
with a $\times 20$ objective and it is 
recorded with an u-eye 2240c camera at a rate of 10 frames per second (fps). In 
order to capture only the motion near the bottom wall or the top wall, 
respectively, the focal plane of the microscope is adjusted 
correspondingly and only the motion of those particles that remain 
in focus is used to determine the self-propulsion velocity. The latter is 
obtained by following Ref. \citenum{Howse2007}: from the tracked two-dimensional 
trajectories of the particles along the wall, as shown in, c.f., Figs. 
\ref{fig:tracked_traj_PS} and \ref{fig:tracked_traj_Si}, the mean square 
displacement as a function of time is determined. The velocity $U$ is 
obtained by fitting the time dependence with the active Brownian motion 
model introduced in Ref. \citenum{Howse2007}. In these calculations, we have 
employed between 40 to 60 tracked trajectories, for each particle type, size, 
and location (at the floor or, if occurring, at the ceiling) of the sliding 
state. For Pt/PS particles, we obtain $U_{PS} \approx 1.5 \pm 0.4~\mathrm{\mu 
m/s}$, for particles of both sizes and for sliding states at both the floor 
and the ceiling walls (see, c.f., the section on results and 
discussion); for Pt/SiO$_2$ particles sliding at the floor (no sliding 
along the ceiling has been observed), the values obtained depend on the size of 
the particle: $U_{Si} \approx 2.1 \pm 0.6~\mathrm{\mu m/s}$ for the small 
($R = 0.5~\mathrm{\mu m}$) particles, and $ \approx 1.4 \pm 0.4~\mathrm{\mu 
m/s}$ for the large ($R = 2~\mathrm{\mu m}$) ones. These velocities compare 
well with simple estimates of the average velocity along the quasi-straight 
parts of the trajectories (see also the videos SI.V1 - SI.V7 in the 
Supporting Information (SI)). 

\section{Model and theory \label{theory}}

As noted in the Introduction, an important question addressed in the 
present study is whether the phenomenology of the emergence of sliding states 
at the ceiling for chemically active, gyrotactic particles can be captured by 
the same simple models, which have been employed in the previous 
investigations of the behavior at the floor 
\cite{Uspal2015a,Koplik2016,Uspal2018b,Uspal2018c,Liverpool2017}. 
Accordingly, here we use the framework of self-phoresis and, concerning the 
model chemical activity, we study both the ``constant flux'' model, employed in 
previous studies \cite{Golestanian2007,Uspal2015a,Koplik2016}, as well as 
a model with spatially ``variable  flux'' across the active cap. The 
latter model allows for a possible dependence of the catalytic 
activity on the local thickness of the catalytic layer. This layer is 
assumed to vary from being thick at the pole to being (vanishingly) 
thin at the equator, as proposed in Refs. 
\citenum{Ebbens2013} and \citenum{Ebbens2019} (for more details see below 
and, 
c.f., Fig. \ref{fig4}).

In brief, the chemical activity of the Janus colloid is modeled as the 
release, at the catalytic cap, of a molecular species into the surrounding 
solution. (Here it is $\mathrm{O_2}$ 
resulting from the Pt-catalyzed decomposition of $\mathrm{H_2 O_2}$; 
the other product of the reaction is the solvent $\mathrm{H_2 O}$.) The 
reaction is assumed to be within the regime of reaction-limited kinetics, 
and $\mathrm{H_2 O_2}$ is taken to be present in abundance. This case is 
likely to apply for typical experiments, with very dilute particle suspensions, 
large volumes of solution with few percent v/v concentration of 
$\mathrm{H_2 O_2}$, and typical duration of experiments of the order of 
less then 30 min. 

%%%%%%%%%%%%%%%%%%%%%%%%%%%%%%%%%%%%
\begin{figure}[!htb]
    \centering
    \includegraphics[width=.95\columnwidth]{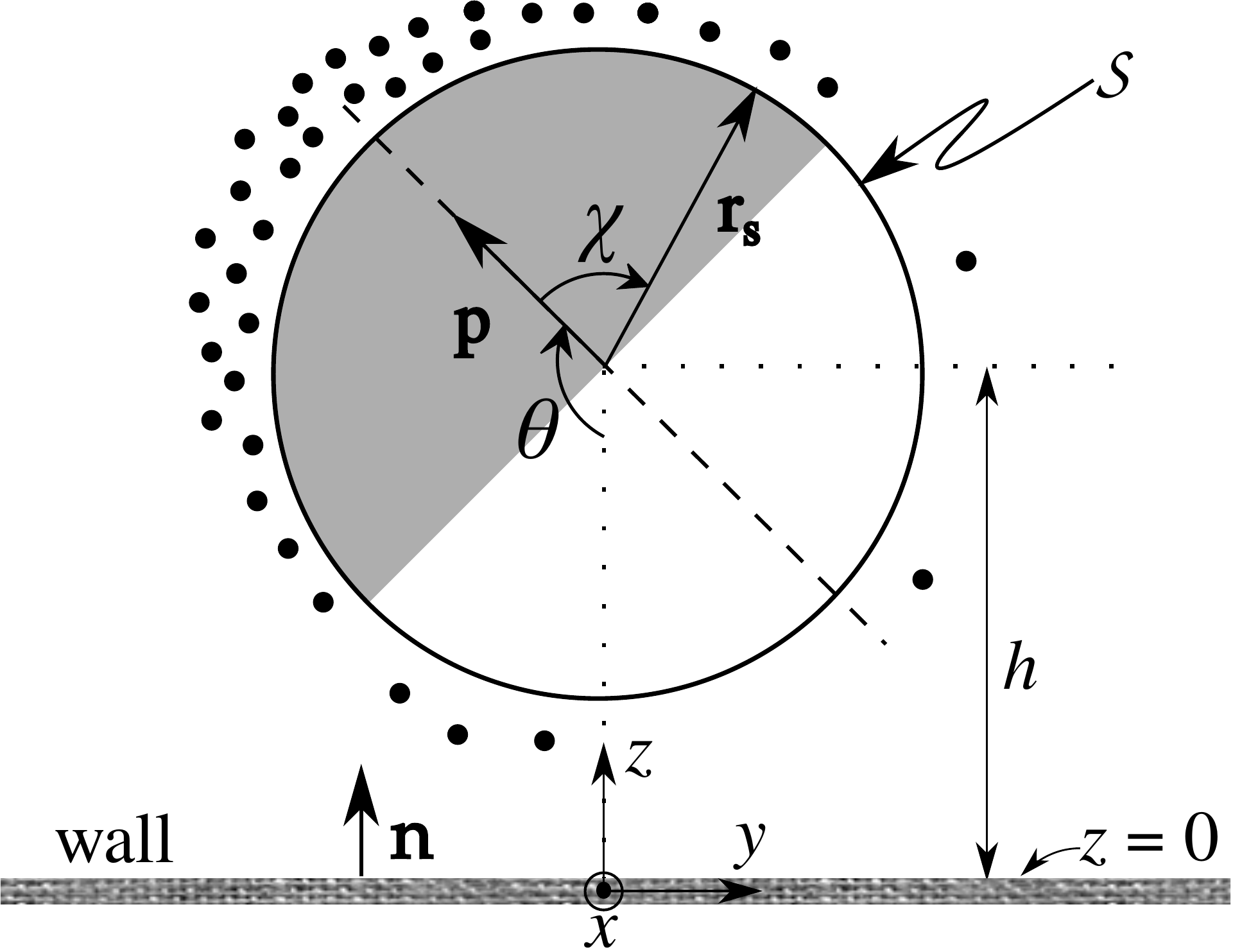}
     \caption{
\label{fig3}
{\small 
A spherical, chemically active Janus particle of radius $R$, with its 
center at (dimensionless) height $h = {\tilde h}/R$ above a planar wall located 
at $z = 0$ and with orientation $\theta$ defined by $\cos \theta = - \mathbf{p} 
\cdot \mathbf{e}_z$; the pair $(h, \theta)$ is referred to as 
``configuration''. 
The active (inert) part is indicated by dark gray (white) color. The location 
$\mathbf{r}_s$ on the surface $\mathcal{S}$ of the particle is parameterized by 
the angle $\chi$ between $\mathbf{r}_s$ and $\mathbf{p}$. The inner normal of 
the wall is shown by the unit vector $\mathbf{n}$. All lengths without 
tilde are measured in units of $R$, and thus they are dimensionless.
}
}
\end{figure}
%%%%%%%%%%%%%%%%%%%%%%%%%%%%%%%%%%%%
The solute molecules diffuse in the surrounding solution with diffusion 
coefficient $D$; the diffusion of the solute is assumed to be a very fast 
process compared to both the motion of the particle and the advection of the 
solute by the flow of the solution; i.e., for the solute transport P{\'e}clet 
number one has $\mathrm{Pe} :=  U_0 R / D \ll 1$, where $R$ is the radius of the 
particle, and $U_0$ is a characteristic particle velocity. Under these 
latter assumptions, the solute number density field relaxes quickly to a 
quasi-steady 
distribution $c(\mathbf{r})$. Assuming that $c(\mathbf{r})$ is sufficiently 
small, such that the chemical potential of the solute can be approximated 
by that of an ideal gas, the distribution $c(\mathbf{r})$ is the solution of the 
Laplace equation, 
\begin{equation}
\label{eq:Laplace}
\nabla^2 c(\mathbf{r}) = 0\,,~\mathbf{r} \in~\mathrm{fluid}\,,
\end{equation}
subject to boundary conditions at the surface $\mathcal{S}$ of the particle, at 
the planar wall located at $z = 0$, see Fig. \ref{fig3}, 
and at infinity (far from the particle, i.e., in the ``bulk'' solution). These 
are given by \newline
\begin{subequations}
 \label{eq:BCs_Laplace}
 \begin{equation}
  \label{eq:BC_c_part}
  [\mathbf{n}\cdot (-D \nabla c)]|_{\mathbf{r}_s} = \mathcal{Q} 
q(\mathbf{r}_s)\,,~\mathbf{r}_s \in ~\mathcal{S} \,,
 \end{equation}
 \begin{equation}
  \label{eq:BC_c_wall}
  [\mathbf{n} \cdot (-D \nabla c)]|_{z = 0} = 0 \,,
 \end{equation} 
and 
 \begin{equation}
  \label{eq:BC_c_infty}
  c(|\mathbf{r}| \to \infty)  = 0\,,~ \mathbf{r} \in~\mathrm{fluid}\,.
 \end{equation}
\end{subequations}
Here, the unit vector $\mathbf{n}$ denotes the \textit{inner} 
(i.e., oriented into the fluid) normal of the corresponding boundary (the 
particle surface or the wall). The dimensionless function $q(\mathbf{r}_s)$ 
accounts for the geometrical distribution of the chemical activity across the 
surface. Accordingly, one has $q(\mathbf{r}_s) = 0$ at the locations 
$\mathbf{r}_s$ 
(position vector from the center of the particle, see Fig. \ref{fig3}) within 
the inert part of the surface. In addition, 
$q(\mathbf{r}_s)$ takes into account possible variations 
of the reaction rate across the catalyst-covered, active area (see 
the discussion below). The factor $\mathcal{Q} > 0$ (with units of 
$\mathrm{m^{-2} \times s^{-1}}$) is fixed by noting that the total rate of 
solute production, which is experimentally measurable, equals $\mathcal{Q} 
\int_\mathcal{S} d^2\mathrm{r}_s q(\mathbf{r}_s)$.

The two model chemical activities, which we consider in this study, 
correspond to the following choices for the function $q(\mathbf{r}_s)$ at 
locations $\mathbf{r}_s$ on the surface of the active hemisphere (see Fig. 
\ref{fig3}). 
\begin{subequations}
\label{eq:choices_f}
In the ``\textit{c}onstant \textit{f}lux'' model \cite{Golestanian2007} one 
has 
\begin{equation}
 \label{eq:def_const_flux}
 q_{cf}(\mathbf{r}_s) = 1\,, \mathbf{r}_s \in~\mathrm{catalyst}\,,
\end{equation}
whereas in the ``\textit{v}ariable \textit{f}lux'' model 
\cite{Ebbens2010}, 
\begin{equation}
 \label{eq:def_var_flux}
 q_{vf}(\mathbf{r}_s) = \cos(\chi)\,,~0 \leq \chi \leq \pi/2\,,
\end{equation}
with (see Fig. \ref{fig3}) $\chi$ denoting the angle between $\mathbf{r}_s$ and 
the director $\mathbf{p}$ (which is the unit vector along the symmetry axis of 
the particle, oriented towards the active pole). 
Equation (\ref{eq:def_var_flux}) corresponds to an activity that decreases from 
a  maximum at the catalytic pole to zero at the equator (i.e., the circle 
separating the active and inert hemispheres).
\end{subequations}

For a given configuration, i.e., a dimensionless height $h = \tilde h/R$ of the 
center of the particle above the wall and an orientation $\theta$, defined by 
the angle between $\mathbf{p}$ and the \textit{outer} normal, $\mathbf{n} = 
-\mathbf{e}_z$, of the wall (see Fig. \ref{fig3}), Eqs. (\ref{eq:Laplace}) and 
(\ref{eq:BCs_Laplace}), with either of the choices for $q(\mathbf{r}_s)$ in Eqs. 
(\ref{eq:choices_f}), can be solved numerically by using the Boundary 
Element Method \cite{pozrikidis02} (BEM) for determining the distribution 
of the solute $c(\mathbf{r})$ (see Ref. \citenum{Uspal2015a}).

The coupling of the chemical field to the motion of the particle and to 
the hydrodynamic flow of the solution is described by assuming the framework 
of self-diffusiophoresis with phoretic slip 
\cite{Anderson1989,Golestanian2007}. Accordingly, the short-ranged interaction 
of solute molecules (which are non-uniformly distributed across the surface) 
with the particle drives a flow within a thin layer adjacent to the 
surface. This surface flow, encoding the 
actuation of the surrounding liquid by the active particle, is modeled 
as an effective slip velocity, usually called ``phoretic slip''; i.e., the 
motion of the fluid relative to the surface of the particle is given by
\begin{equation}
 \label{eq:phor_slip}
\mathbf{v}_s(\mathbf{r}_s) = -b(\mathbf{r}_s) \nabla_{||} 
c(\mathbf{r}_s)\,,~~\mathbf{r}_s \in~\mathcal{S}\,.  
\end{equation}
Here, $\nabla_{||} \equiv (\hat{\mathcal{I}} -  \mathbf{n} \mathbf{n}) 
\cdot \nabla$ is the surface gradient operator,  with $\hat{\mathcal{I}}$ 
denoting the identity tensor. The so-called surface mobility $b(\mathbf{r}_s)$ 
encodes the solute-particle interaction. We consider 
\begin{equation}
 \label{eq:def_b}
 b(\mathbf{r}_s) = 
 \begin{cases}
  ~b_c = constant\,,~\mathbf{r}_s 
\in \mathrm{\textit{c}atalyst}\,,\\
  ~b_i  = constant \,,~\mathbf{r}_s \in \mathrm{\textit{i}nert}\,,
 \end{cases}
\end{equation}
because these two regions consist of different materials (Pt and PS or 
$\mathrm{SiO_2}$). In general, a similar mechanism gives rise to osmotic flows 
along the \textit{w}all, characterized by a corresponding surface mobility 
$b_w$; in this work, we restrict the discussion to the case of a no-slip wall 
(i.e., $b_w = 0$).

The active phoretic slip at the surface of the particle drives the hydrodynamic 
flow of the solution and the motion of the particle, with rigid body 
translational and rotational velocities $\mathbf{U}$ and $\mathbf{\Omega}$, 
respectively. For micrometer-sized chemically active particles, moving with 
velocities of the order of micrometer/s through water-like liquids, the induced 
flows are characterized by a very small Reynolds number $\mathrm{Re} \equiv \rho 
U_0 R / \mu \ll 1$, where $R$ is the radius of the particle, and $\mu$ and 
$\rho$ are the viscosity and mass density of the solution, respectively 
\cite{Anderson1989,SenRev,Bechinger2016_rev,Posner2017,Golestanian2007}. 
Accordingly, the flow field $\mathbf{v}(\mathbf{r})$ is governed by the 
incompressible Stokes equations 
\begin{equation}
 \label{eq:Stokes}
 \nabla \cdot \hat{\boldsymbol{\sigma}} = 0\,,~~\nabla 
\cdot \mathbf{v}(\mathbf{r}) = 0\,,
\end{equation}
where the stress tensor $\hat{\boldsymbol{\sigma}}$ is taken to be 
that of a Newtonian fluid, i.e., $\hat{\boldsymbol{\sigma}} = -P 
\hat{\mathcal{I}} + \mu \left[ \nabla \mathbf{v} + (\nabla 
\mathbf{v})^\mathrm{T} \right]$ and $P(\mathbf{r})$ is the fluid pressure 
enforcing the incompressibility. The solution is subject to the boundary 
conditions at the surface of the particle (phoretic slip)
\begin{subequations}
 \label{eq:BCs_Stokes}
 \begin{equation}
  \label{eq:BC_u_part}
  \mathbf{v}(\mathbf{r}_s) = \mathbf{U} + \mathbf{\Omega} \times \mathbf{r}_s 
  + \mathbf{v}_s(\mathbf{r}_s)\,,~\mathbf{r}_s \in~\mathcal{S}\,,
 \end{equation}
at the wall (no slip)
  \begin{equation}
  \label{eq:BC_u_wall}
  \mathbf{v}(\mathbf{r})|_{z = 0} = 0\,,
 \end{equation}
and at infinity (quiescent fluid)
   \begin{equation}
  \label{eq:BC_u_infty}
  \mathbf{v}(|\mathbf{r}| \to \infty) = 0\,.
 \end{equation}
\end{subequations}
The above equations, which depend on the yet unknown quantities $\mathbf{U}$ and 
$\mathbf{\Omega}$, are closed by the conditions of vanishing net force and net 
torque on the particle corresponding to an overdamped motion. These 
conditions render the relations
\begin{subequations}
 \label{eq:F_and_T_bal}
 \begin{equation}
  \label{eq:F_bal}
  \int\limits_\mathcal{S} d^2 \mathrm{r}_s~ 
\,\hat{\boldsymbol{\sigma}}\cdot 
\mathbf{n} + \mathbf{F}_g = 0
 \end{equation}
 and
 \begin{equation}
  \label{eq:T_bal}
  \int\limits_\mathcal{S} d^2 \mathrm{r}_s~ \,\mathbf{r}_s \times 
(\hat{\boldsymbol{\sigma}} \cdot \mathbf{n}) + \mathbf{T}_g = 0\,,
 \end{equation}
\end{subequations}
where the first terms on the left hand sides are the hydrodynamic force and 
torque acting on the particle, while the second terms are the force and 
the torque accounting for the effects of gravity (including buoyancy). The 
latter can be calculated separately, upon knowing the material specifications 
of the Janus particles (core material, catalyst material, and the geometry of 
the layer of catalyst). 

With $c(\mathbf{r})$ known, the phoretic slip is determined, and Eqs. 
(\ref{eq:BCs_Stokes}) - (\ref{eq:F_and_T_bal}) can be solved numerically by 
using BEM to calculate, for each configuration $(h,\theta)$ of interest, 
the velocities $\mathbf{U}$ and $\mathbf{\Omega}$, as well as the flow field 
$\mathbf{v}(\mathbf{r})$ (see Refs. \citenum{Uspal2015a} and 
\citenum{Simmchen2016}). The calculation can be simplified by exploiting the 
linearity of the Stokes equations as follows. First, the velocities can be 
written as the superposition of a phoretic  (index ``ph'') component and a 
gravity-induced one (index ``g''),
\begin{equation}
 \label{eq:vel_superp}
 \mathbf{U} = \mathbf{U}_{ph} + \mathbf{U}_{g}\,,~ 
 \mathbf{\Omega} = \mathbf{\Omega}_{ph} +\mathbf{\Omega}_{g}\,.
\end{equation}
The phoretic component is obtained by solving Eqs. (\ref{eq:BCs_Stokes}) - 
(\ref{eq:F_and_T_bal}) without external forces and torques, i.e., with  
$\mathbf{F}_g = 0$ and $\mathbf{T}_g =0$, while the gravity-induced component 
is obtained by solving Eqs. (\ref{eq:BCs_Stokes}) - (\ref{eq:F_and_T_bal}) 
without phoretic slip, i.e., with $\mathbf{v}_s(\mathbf{r}_s) \equiv 0$ in Eq. 
(\ref{eq:BC_u_part}).  Furthermore, the linearity of the Stokes equations 
also implies that the phoretic velocity of a Janus particle, with phoretic 
mobilities  $(b_c,b_i)$ across the two parts of the surface, can be 
expressed in terms of the phoretic velocities corresponding to the pairs of 
phoretic mobilities $(1,0)$ and $(0,1)$ as \cite{Bayati2019}
\begin{eqnarray}
 \label{eq:phor_vel_superp}
 \mathbf{U}_{ph} &=& b_c \mathbf{U}_{ph}^{(1,0)} + 
b_i \mathbf{U}_{ph}^{(0,1)}\,, \nonumber\\ 
 \mathbf{\Omega}_{ph} &=& b_c \mathbf{\Omega}_{ph}^{(1,0)} 
+ b_i \mathbf{\Omega}_{ph}^{(0,1)}\,.
\end{eqnarray}
The  linearity of the Stokes equations also implies that the velocities 
induced by the force and torque due to gravity can be written as
\begin{equation}
 \label{eq:mob_matrix_connect}
 \begin{pmatrix}
\mathbf{U}_{g} \hfill \\
\mathbf{\Omega}_{g} \hfill
\end{pmatrix} = \mathbb{M} 
 \begin{pmatrix}
\mathbf{F}_g \hfill \\
\mathbf{T}_g \hfill
\end{pmatrix}\,,
\end{equation}
where the $6 \times 6$ mobility matrix $\mathbb{M}$ depends solely on the 
geometry of the system \cite{HaBr73}; i.e., in the present case, $\mathbb{M}$ 
depends only on the spherical shape of the particle and on the dimensionless 
distance $h$ of its center from the wall. In summary, from Eqs. 
(\ref{eq:vel_superp})-(\ref{eq:mob_matrix_connect}) one concludes that for our 
model system it is sufficient to solve, for a given configuration, only 
the two phoretic problems $(b_c,b_i) = (1,0)$ and $(0,1)$ and the classical 
hydrodynamic mobility problem in order to determine $\mathbb{M}$. (For a 
sphere near a wall, the sparse structure of $\mathbb{M}$, with several 
vanishing entries, is a well known result in the low Reynolds number literature 
\cite{HaBr73}). Knowledge of the three quantities mentioned above allows 
one to determine the translational and rotational velocities of the Janus 
particle for any phoretic mobility $(b_c,b_i)$, subject to any constant external 
forces $\mathbf{F}_g$ and torques $\mathbf{T}_g$. This provides a huge reduction 
of the computational costs of the study.

We now turn to the issue of the motion of the Janus particle. This motion is 
calculated as follows. With a focus on understanding the deterministic 
dynamics of the particle, in particular the steady states of the deterministic 
dynamics, here we disregard the influence of thermal fluctuations on the motion 
of the particle. As discussed in Ref. \citenum{Uspal2015a}, and studied in 
detail in Ref. \citenum{Koplik2016}, the sliding steady states of the 
deterministic dynamics turn out to be, in general, quite robust with 
respect to thermal fluctuations.  Accordingly, in the fixed laboratory frame of 
reference the overdamped motion of the particle (i.e., translation of the 
center of mass $\mathbf{r}_O$ and rotation of $\mathbf{p}$) follows from the 
velocities  $\mathbf{U}$ and $\mathbf{\Omega}$ introduced above:
\begin{subequations}
 \label{eq:part_mot_gen}
 \begin{equation}
  \label{eq:CM_trans_gen}
  \frac{d \mathbf{r}_O}{dt } = \mathbf{U}(h,\theta)
 \end{equation}
and
\begin{equation}
  \label{eq:rot_p_gen}
  \frac{d\mathbf{p}}{dt} = \mathbf{\Omega}(h,\theta) \times \mathbf{p} \,,
\end{equation}
 \end{subequations}
where $t$ denotes time. (The dependences of the velocities on position and 
orientation have been explicitly indicated here.) The axial symmetry of 
the Janus particle implies that the phoretic motion problem, i.e., the motion 
in the absence of external forces and torques, possesses reflection symmetry 
with respect to the plane that is normal to the wall and contains the 
director $\mathbf{p}$. Moreover, the forces due to gravity, i.e., weight and 
buoyancy, lie also in this plane, while the torque due to gravity points 
into the direction normal to this plane. Consequently, one infers that the 
deterministic motion of the particle is two-dimensional, i.e., it is confined 
to that plane normal to the wall, which contains the initial orientation 
(at $t = 0$) of the particle. Accordingly, we can choose the 
system of coordinates (see Fig. \ref{fig3}) such that the trajectory of the 
particle lies within the $z-y$ plane.

The equations of motion are rendered dimensionless as follows. From the phoretic 
motion problem, one can infer a characteristic translational velocity $U_0 := 
|b_c| \mathcal{Q}/D$ and, accordingly, a characteristic rotational velocity 
$\Omega_0 := U_0/R$ as well as a characteristic time $t_0 := 1/\Omega_0$. 
In terms of 
the characteristic velocity $U_0$, the Stokes formula for the drag on a sphere 
in unbounded space gives a characteristic force $F_0 := 6 \pi \mu R U_0$, which 
provides the characteristic torque $T_0 := F_0 R = 6 \pi \mu R^2 U_0$. Defining 
\begin{equation}
 \label{eq:def_beta} 
 \beta := b_i/b_c\,,
\end{equation}
we note that the phoretic velocity $\mathbf{U}_{ph}^{(fs)}$, which such Janus 
particles would exhibit in an unbounded solution (``\textit{f}ree 
\textit{s}pace''), is related to $U_0$ via
\begin{equation}
 \label{eq:Ufs}
\mathbf{U}_{ph}^{(fs)} = \frac{b_c}{|b_c|} \,\mathcal{U}(\beta) \, U_0 
\,\mathbf{p}\,,
\end{equation}
where the prefactor $\mathcal{U}(\beta)$, which depends on the model for the 
activity, can be calculated analytically \cite{Michelin2014,Popescu2018b}. 
For our choices of the models for the chemical activity, it is given by 
\begin{subequations}
 \label{eq:Ufs_models}
 \begin{equation}
  \label{eq:Ufs_const_flux}
  \mathcal{U}(\beta) = (1+\beta)/8\,,~\mathrm{constant~flux}\,,
 \end{equation}
and
  \begin{equation}
  \label{eq:Ufs_var_flux}
  \mathcal{U}(\beta) \approx 0.129 + 0.038\,\beta\,,~\mathrm{variable~flux}\,.
 \end{equation}
\end{subequations}
(There is no equivalent rotational velocity, because in the 
absence of boundaries the motion of the axisymmetric particle consists only of 
translation along the symmetry axis.) This will prove to be useful for 
obtaining estimates, 
in terms of experimentally measured quantities, for the relevant parameters of 
the dynamics of the active particle (see, c.f., the section on results and 
discussion).    

The dimensionless velocities $\mathbf{u}$ and $\boldsymbol{\omega}$, the 
matrix-elements $m_{i,j}$ of the hydrodynamic mobility $\mathbb{M}$, and the 
gravity-induced forces and torques (including buoyancy) are introduced by the 
following definitions:
\begin{subequations}
 \label{eq:def_dimensionless_quant}
\begin{equation}
\label{eq:scaled_us}
\mathbf{U}^{(1,0)}_{ph} =: U_0 \mathbf{u}^{(1,0)}\,,~
\mathbf{U}^{(0,1)}_{ph} =: U_0 \mathbf{u}^{(0,1)}\,
\end{equation}
\begin{equation}
\label{eq:scaled_oms}
\mathbf{\Omega}^{(1,0)}_{ph} =: \Omega_0 \boldsymbol{\omega}^{(1,0)}\,,~
\mathbf{\Omega}^{(0,1)}_{ph} =: \Omega_0 \boldsymbol{\omega}^{(0,1)}\,,
\end{equation}
\begin{eqnarray}
\label{eq:scaled_Ms}
\mathbb{M}_{3,3} &=:& \frac{m_{3,3}}{6 \pi \mu R} \,,~
\mathbb{M}_{4,4} =: \frac{m_{4,4}}{6 \pi \mu R^3}\,,~\nonumber\\
\mathbb{M}_{2,4} &=: & \frac{m_{2,4}}{6 \pi \mu R^2}\,,
\end{eqnarray}
\begin{equation}
\label{eq:scaled_Fg}
 \mathbf{F}_{g} =: \pm \,F_0 \,|\mathcal{U}(\beta)| \,\mathcal{F}\, 
\mathbf{e}_z\,,
\end{equation}
and 
\begin{equation}
\label{eq:scaled_Tg}
 \mathbf{T}_{g} =: \mp \,T_0 \,|\mathcal{U}(\beta)|\, \mathcal{T} \sin 
\theta \,\mathbf{e}_x\,,
\end{equation}
where $\mathcal{F} > 0$ and $\mathcal{T} > 0$. The $+$ and $-$ signs correspond 
to the situations at the ceiling ($\mathbf{g} = g \mathbf{e_z}$, 
see Figs. \ref{fig1}(left) and \ref{fig3}) and to the situation at the floor 
($\mathbf{g} = - g \mathbf{e_z}$, see Figs. \ref{fig1}(right) and \ref{fig3}), 
respectively.
\end{subequations}
In terms of these dimensionless variables, and by introducing the dimensionless 
lateral position $y: = \tilde{y}/R$ (similarly to $h := 
\tilde{h}/R$, see Fig. \ref{fig3} and below Eq. (\ref{eq:def_var_flux})), the 
equations of motion for the active particle take the form
\begin{eqnarray}
 \label{eq:dimless_motion_part}
 \frac{dh}{d\tau} &=& \frac{b_c}{|b_c|} \left[u_z^{(1,0)}(h,\theta) + 
 \beta u_z^{(0,1)}(h,\theta)\right] \nonumber\\
 &\pm& m_{3,3}(h,\theta) |\mathcal{U}(\beta)| \mathcal{F}\,,\nonumber\\ 
&&\nonumber\\
  \frac{d\theta}{d\tau} &=& -\frac{b_c}{|b_c|} \left[\omega_x^{(1,0)}(h,\theta) 
+  \beta \omega_x^{(0,1)}(h,\theta)\right] \nonumber\\
 &\pm& m_{4,4}(h,\theta) |\mathcal{U}(\beta)| \mathcal{T} 
\sin\theta\,,\\
&&\nonumber\\
 \frac{dy}{d\tau} &=& \frac{b_c}{|b_c|} \left[u_y^{(1,0)}(h,\theta) + 
 \beta u_y^{(0,1)}(h,\theta)\right] \nonumber\\
 &\mp& m_{2,4}(h,\theta)  |\mathcal{U}(\beta)| \mathcal{T} \sin 
\theta\,,\nonumber
\end{eqnarray}
where $\tau$ is a dimensionless time, i.e., it is expressed in 
units of $t_0$ (see above Eq. (\ref{eq:def_beta})).

As anticipated, in light of previous studies 
\cite{Uspal2015a,Simmchen2016,Uspal2018c}, the first two equations are decoupled 
from the third, and thus one can focus on studying the reduced dynamics in the 
$(h,\theta)$ plane. The form of the equations above is also very transparent 
in what concerns the meaning of the dimensionless parameters $\mathcal{F}$ and  
$\mathcal{T}$: they characterize the importance of the gravity-induced effects, 
relative to those due to the active motility. Consequently, if 
$\mathcal{F}$ and  $\mathcal{T}$ are very small, the dynamics is dominated by 
the active motility and one expects to recover the results of the 
corresponding previous studies \cite{Uspal2015a}, while for very large 
values of $\mathcal{F}$ and  
$\mathcal{T}$ the ensuing dynamics should resemble that of sedimentation of a 
heavy, gyrotactic, chemically inert sphere (see, e.g., Ref. 
\citenum{Wirth2019}).

The dynamics of the particle (Eq. (\ref{eq:dimless_motion_part})) depends on 
four particle-dependent material parameters: $\beta$, 
$\mathcal{F}$, $\mathcal{T}$, and the sign of $b_c$. The description of the 
model is completed by providing the geometry of the catalyst, i.e., the active 
layer, which allows one to determine the gravity-induced quantities 
$\mathcal{F}$ and $\mathcal{T}$. Here, for the Janus particle we use the model 
of an ``egg-shell'', i.e., a slightly-deformed spherical shape, as proposed in 
Ref. \citenum{Ebbens2013}. According to this model, the catalyst layer is 
taken to occupy the volume between a sphere of radius $R$ and half of a 
concentric prolate spheroid of minor semi-axis $R$ and major semi-axis 
$R+\delta$, with $\delta/R \ll 1$ (see Fig. \ref{fig4}). 
%%%%%%%%%%%%%%%%%%%%%%%%%%%%%%%%%%%%%%
\begin{figure}[!tb]
    \centering
    \includegraphics[width=.75\columnwidth]{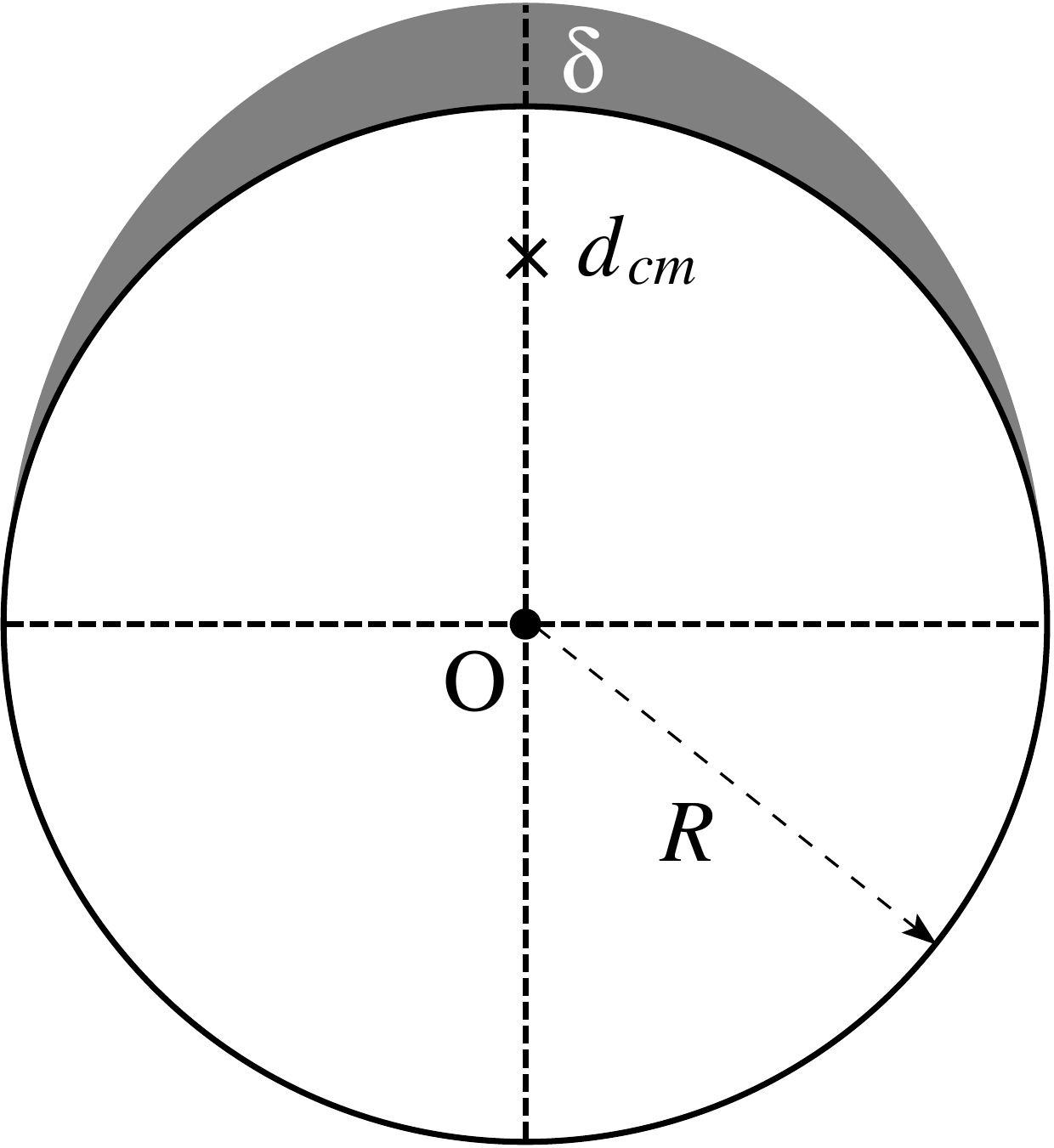}
     \caption{
\label{fig4}
{\small 
Schematic cross-section (not to scale) of the ``egg-shell'' model for a Janus 
particle. The white disk of radius $R$ depicts the spherical core of 
catalytically inactive material, while the lens-like region (dark gray area) 
depicts the catalyst; $\delta$ is the maximum thickness of the catalyst layer 
and $d_{cm}$ is its center of mass.
}
}
\end{figure}
%%%%%%%%%%%%%%%%%%%%%%%%%%%%%%%%%%%%%%
Accordingly, the mass of the catalyst, i.e., of the \textit{a}ctive 
material, within the lens-shaped layer is given by
\begin{equation}
\label{eq:def_delta}
 m_{a} = \frac{2 \pi}{3} \delta R^2 \rho_a\,,
\end{equation}
where $\rho_a$ is the mass density of the catalyst. The \textit{c}enter of 
\textit{m}ass of the layer is located on the symmetry axis at a distance $d_{cm} 
= \frac{3}{4} \left(R+ \frac{1}{2}\, \delta\right) \approx \frac{3}{4} R$ from 
the center O of the sphere \cite{Ebbens2013}. Disregarding the very small 
deviation from  the spherical shape (which concerns the occurrence of 
a torque due to buoyancy) as well as the very small displacement of the center 
of mass of the particle, i.e., core plus shell, from the center O of the 
sphere, the gravity-induced rotation of the sphere relative to its center O is 
approximated by that due to the weight of the catalyst layer applied at  
$d_{cm}$. Therefore, the dimensionless force 
$\mathcal{F}$ and torque $\mathcal{T}$ (Eqs. (\ref{eq:scaled_Fg}) and 
(\ref{eq:scaled_Tg})) are given by
\begin{subequations}
\label{eq:defs_F_T}
 \begin{equation}
  \label{param_F}
\mathcal{F} = \frac{1}{9} \,\frac{2 R^2 (\rho_s - \rho) g}{\mu \,
|\mathbf{U}_{ph}^{(fs)}|} 
\left[1 + \frac{\delta}{2 R}\frac{\rho_a - \rho}{\rho_s - \rho} \right] 
=: 
\frac{1}{9} F\,, 
 \end{equation}
and
 \begin{equation}
  \label{param_T}
 \mathcal{T} = \frac{1}{12} \frac{\rho_a g}{\mu \, |\mathbf{U}_{ph}^{(fs)}|} R 
\delta =: \frac{1}{12} T\,,
 \end{equation}
\end{subequations}
where $\rho_s$ denotes the mass density of the spherical core material. We 
emphasize that $T$ depends on geometrical parameters and on the density of 
the active material (catalyst), but not on the density of the spherical core. 
Accordingly, as noted in the Introduction, the torque due to gravity, i.e., the 
gyrotactic response of the particle, is generically relevant for Janus 
particles, irrespective of the kind of material the core is 
made of.

With these quantities, the model for chemically active, gyrotactic 
(bottom heavy) Janus particles near a planar, no-slip wall is complete, 
and we can proceed to study the dynamics.

\section{Results and discussion \label{results}}

The trajectory of a particle, starting from an initial configuration $h(\tau = 
0) = h_0$ and $\theta(\tau = 0) = \theta_0$, 
is calculated by integrating the equations of motion in Eq. 
(\ref{eq:dimless_motion_part}), using a procedure which follows closely 
the methodology used in Refs. \citenum{Uspal2015a,Simmchen2016}, and 
\citenum{Bayati2019}. The 
functions $\mathbf{u}^{(1,0)}(h,\theta)$, $\mathbf{u}^{(0,1)}(h,\theta)$, 
$\boldsymbol{\omega}^{(1,0)}(h,\theta)$, 
$\boldsymbol{\omega}^{(0,1)}(h,\theta)$, and $\mathbb{M}(h)$ are evaluated 
on a grid of values $h_i$ spanning the range $1.02 \leq h \leq 10$ 
and (at each $h_i$) a grid of values $\theta_j$ spanning the range $0 \leq 
\theta \leq \pi$. The results are stored as five tables of values. 
(As noted in the section on the model and theory, this calculation has 
to be 
done only once.) The grid $\{h_i\}$ is nonuniform: a dense grid (small step 
$\Delta h = 0.01$) is used for $1.02 \lesssim h \lesssim 1.2$, and then the 
density of evaluation points is gradually decreased towards $\Delta h = 
0.1$ for $h > 5$. The lower cut-off is due to (i) the assumptions 
underlying the model, e.g., that of a boundary layer (mapped onto the phoretic 
slip) being much thinner than the other length-scales in the problem, 
cease to hold; and (ii) for very small values of $h$, the 
computational cost of an accurate calculation of the translational and 
rotational velocities 
by the BEM becomes prohibitive. Accordingly, trajectories which reach 
points with $h \leq 1.02$ are classified as ``crashing-into-the-wall'' 
events. For a set of parameters $(F,T,\beta)$ and for a choice of the model 
activity (i.e., the expression $\mathcal{U}(\beta)$), the corresponding linear 
combination of the stored five tables (see Eq. 
(\ref{eq:dimless_motion_part})) is 
constructed; the right hand side (rhs) of Eq. 
(\ref{eq:dimless_motion_part}) can then be evaluated for any pair 
$(h,\theta)$ by interpolation, and for any initial condition $1.02 < h_0 < 10$ 
and $0 \leq \theta_{0} \leq \pi$ the trajectory $\{h(\tau),\theta(\tau)\}$ 
follows upon numerical integration by using the Euler scheme with 
suitably small (near the wall) or large (far from the wall) time steps. 
In the space $(h,\theta)$ the dynamics is most conveniently 
visualized in the form of phase portraits, i.e., generalized flow fields,  
of the trajectories, starting from various initial conditions $(h_0,
\theta_0)$. 

\subsection{\large Phase portraits of dynamics}
%%%%%%%%%%%%%%%%%%%%%%%%%%%%%%%%%%%%%%
\begin{figure}[!ht]
    \centering
\includegraphics[width=.9\columnwidth]{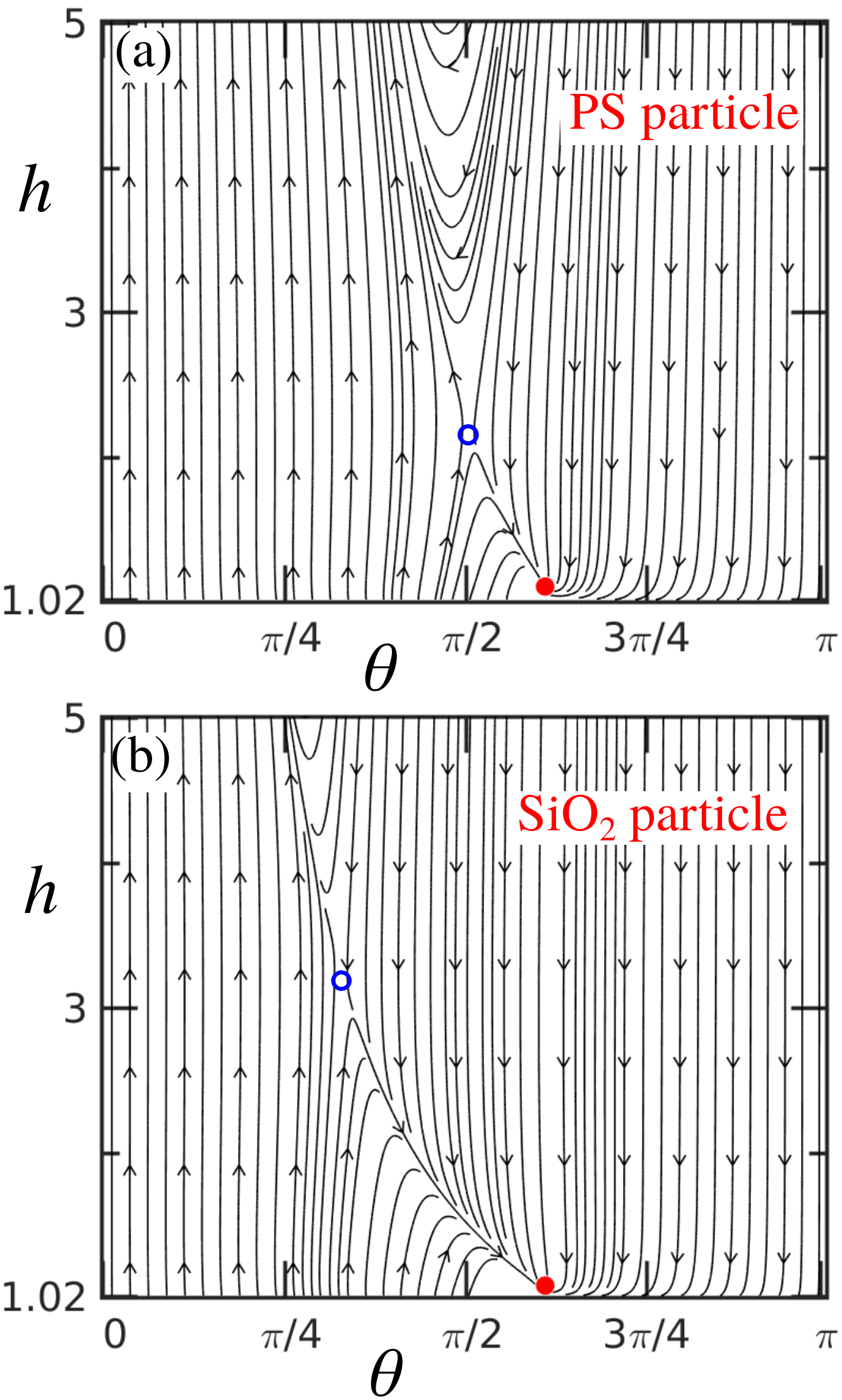}
     \caption{
\label{fig5}
{\small Phase portraits illustrating the dynamics of a chemically active 
Janus particle in the vicinity of the floor. The results correspond to the 
model activity ``constant flux'', $b_c < 0$, with the parameters 
$(F,T,\beta)$ given by (a) (1.2,0.62,0.5) and (b) (7.6,0.31,0.5); 
the red dots indicate the location of the corresponding ``sliding state'' 
attractor, while the open blue symbols indicate the location of an 
unstable, stationary saddle point. The first triplet could correspond to, 
e.g., a spherical Janus particle of radius $R = 
1.5~\mathrm{\mu m}$, made out of a PS core and a Pt film of $\delta = 
7~\mathrm{nm}$, moving with 
$|\mathbf{U}_{ph}^{(fs)}| = 4~\mathrm{\mu m/s}$ (see Eqs. 
(\ref{eq:defs_F_T}) and the discussion below); the second may be the case of, 
e.g., a spherical Janus particle of radius $R = 1.5~\mathrm{\mu m}$, made out of 
a SiO$_2$ core and a Pt film of $\delta = 7~\mathrm{nm}$, moving with 
$|\mathbf{U}_{ph}^{(fs)}| = 8~\mathrm{\mu m/s}$. Note that only the region $h 
\leq 5$ is shown, both for reasons of clarity and because for $h > 5$ the 
dynamics is basically that of a particle in the bulk.
}
}
\end{figure}
%%%%%%%%%%%%%%%%%%%%%%%%%%%%%%%%%%%%%%
An example of such phase portraits is shown in Fig. \ref{fig5} for Janus 
particles with the model chemical activity ``constant flux'', $b_c < 0$, 
and distinct triplets $(F,T,\beta)$ near the ``floor'' (i.e., for 
$\mathbf{g} \cdot \mathbf{e}_z < 0$). In both cases, for the dynamics one 
finds an attractor fixed point $(h^*,\theta^*)$, marked by the red dots 
in Figs. \ref{fig5}(a) and (b). The trajectories converge towards the attractor 
fixed points of the dynamics, i.e., $dh/d\tau = 0$ and $d 
\theta/d\tau = 0$ at $(h^*,\theta^*)$. The simultaneous vanishing of those 
derivatives tells that the height of the particle and its orientation with 
respect to the wall become constant in time; accordingly, the rhs of the third 
part of Eq. (\ref{eq:dimless_motion_part}) becomes constant, too, i.e., the 
$y$-component of 
the velocity turns time-independent. Therefore, these are 
identified as steady states of sliding along the wall. The phase portraits 
also illustrate  all the other generic types of trajectories observed in 
the study of the dynamics of a Janus particle: ``crashing-into-the-wall'' (see, 
e.g., the trajectories approaching the wall in the region $\theta > 3 \pi/4$), 
``reflection'' from the wall (see, e.g., the trajectories at the center top of 
Fig. \ref{fig5}(a)), as well as ``escape'' from the wall (e.g., the trajectories 
in the region $\theta < \pi/4$).

We recall that for small magnitudes of $F$ and $T$ (Eq. \ref{eq:defs_F_T}) 
the dynamics is expected to be similar to that exhibited by the model particle 
in a gravitationless environment. However, in the case $(F,T) = (0,0)$ the 
distinction between floor and ceiling disappears. Thus, if sliding states 
occur 
at one wall, they should occur also at the other one. If this happens, the 
sliding states should also occur, simultaneously, for $(F,T)$ being close 
to $(0,0)$, although the topologies of the corresponding phase portraits 
will start to differ as the location $(F,T)$ moves further away from $(0,0)$. 
The 
expectation of the simultaneous occurrence of sliding states at the 
floor and at the ceiling 
for small values of $F$ and $T$ can be directly tested in the 
context of the same 
model with ``constant flux'' chemical activity and $b_c < 0$. For this 
case it is 
known that sliding states, solely due to phoresis, may occur \cite{Uspal2015a}.
%%%%%%%%%%%%%%%%%%%%%%%%%%%%%%%%%%%%%%
\begin{figure}[!th]
    \centering
\includegraphics[width=.9\columnwidth]{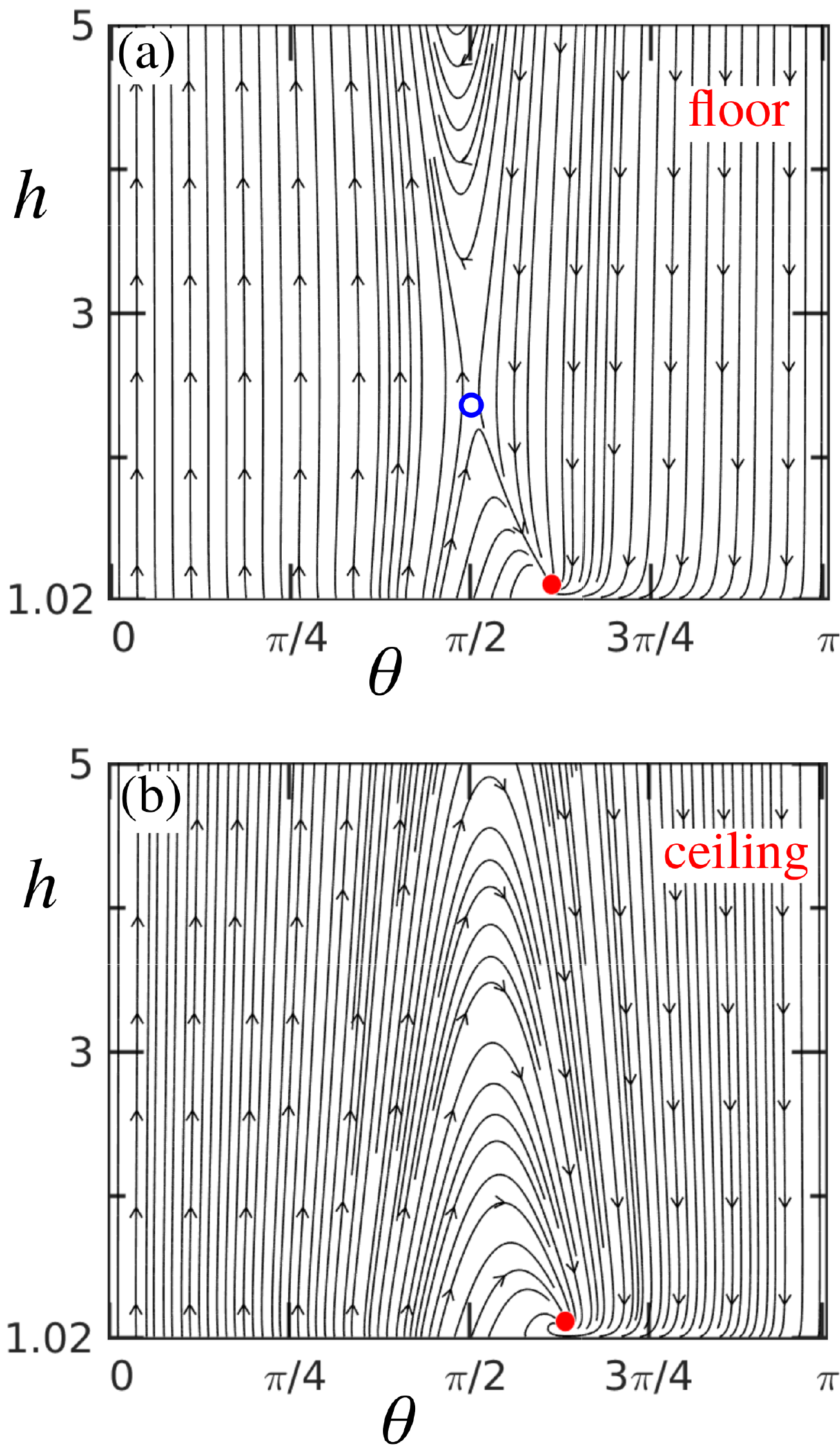}
     \caption{
\label{fig6}
{\small Phase portraits for a model Janus particle with ``constant 
flux'' chemical activity, $b_c < 0$, and the parameters $(F,T,\beta) = 
(0.86,0.49,0.5)$ in the vicinity of (a) the floor and  (b) the 
ceiling; the red dots indicate the location of the corresponding ``sliding 
state'' attractor, while the open blue symbol in panel (a) indicates the 
location of a saddle point. Note that only the region 
$h \leq 5$ is shown, both for 
reasons of clarity and because for $h > 5$ the dynamics is basically that of a 
particle in the bulk.
}
}
\end{figure}
%%%%%%%%%%%%%%%%%%%%%%%%%%%%%%%%%%%%%%

As shown in Fig. \ref{fig6}, this is indeed the case: for small, 
\textit{but non-zero}, $F = 0.86$ and $T= 0.49$, sliding states are 
possible at both walls for \textit{the same value} of $\beta$, i.e., the exact 
same particle and with the same activity, i.e., the same value of 
$U_{ph}^{(fs)}$. Although the corresponding configurations (height $h^*$ 
and orientation $\theta^*$) of the sliding states at the floor and 
at the ceiling are very similar, the topologies of the phase 
portraits are clearly distinct. Moreover, their corresponding 
basins of attractions, i.e., the set of initial conditions for which the 
trajectories converge to the fixed point, have different extents.

%%%%%%%%%%%%%%%%%%%%%%%%%%%%%%%%%%%%%%
\begin{figure}[!h]
    \centering
\includegraphics[width=.9\columnwidth]{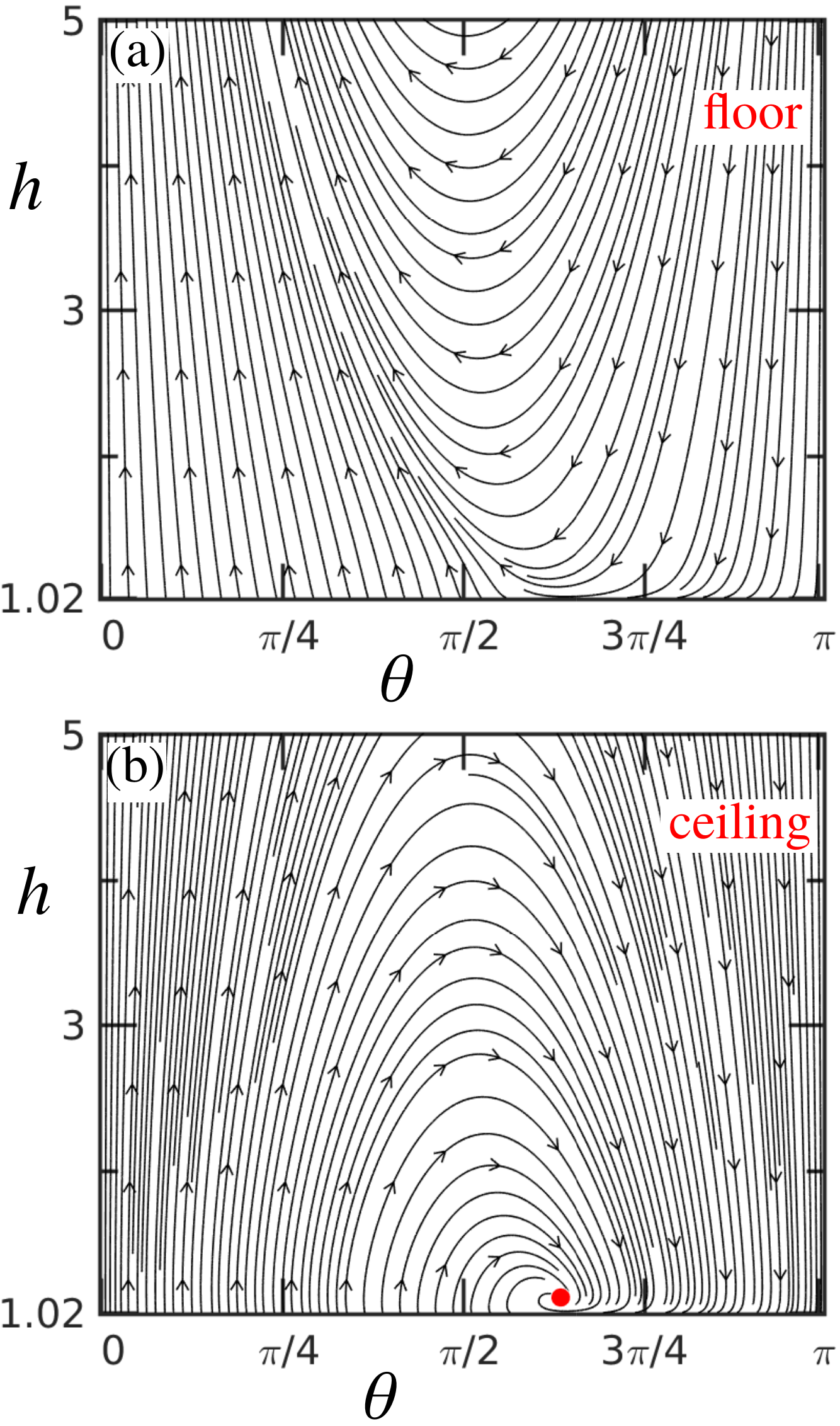}
     \caption{
\label{fig7}
{\small Phase portraits for a model Janus particle with ``constant 
flux'' chemical activity, $b_c < 0$, and the parameters $(F,T,\beta) = 
(0.1,4,0.9)$ in the vicinity of (a) the floor and  (b) the ceiling; 
the red dot indicates the location of the corresponding ``sliding state'' 
attractor. Note that only the region $h \leq 5$ is shown, both for reasons 
of clarity and because for $h > 5$ the dynamics is basically that of a particle 
in the bulk.
}
}
\end{figure}
%%%%%%%%%%%%%%%%%%%%%%%%%%%%%%%%%%%%%%
As noted in the Introduction, another scenario reported in experimental studies 
is that Janus particles, which -- owing to a sufficiently large 
chemical activity -- move upwards (against gravity, eventually after a lift-off 
from the floor), upon collision with the ceiling attain a sliding 
state along it \cite{Ebbens2013,Ebbens2019,Brown2014,Takatori2016}. As 
shown in Fig. \ref{fig7}, this scenario can be captured also by the model 
Janus particle with a ``constant flux'' chemical activity and $b_c < 0$.

\subsection{\large Theoretical state-diagrams}
The results discussed above highlight the rich behavior exhibited by the 
dynamics of the model Janus particle, as well as the fact that the model 
captures (qualitatively) the types of sliding states and the set-up (floor 
or ceiling) encountered in experimental studies.\bibnote{Additionally, 
for suitable choices of $(F,T,\beta)$, for both models the dynamics exhibits 
also the previously reported ``hovering'' states, both at the floor and 
the ceiling. 
These states have proven to be difficult to detect experimentally, 
with 
only one report of observation seemingly compatible with such a state 
\cite{Uspal2018c}. We only succinctly discuss them in the SI Sec. 
\ref{sec:appB}, 
and focus the discussion here on the sliding states.} Moreover, the feature of 
co-existing sliding states at the floor and the ceiling 
\cite{Brown2014,Howse2015} is also predicted to occur for certain values of the 
parameters, too (see Fig. \ref{fig6}). (We note that, as discussed 
below, 
similar results can be obtained by using the ``variable flux'' model with $b_c < 
0$, but, obviously, for other values of the triplet $(F,T,\beta)$.) It is 
therefore important to understand how the occurrence of the (experimentally 
observable) sliding states at the ceiling or at the floor (or both, or 
none) depends on the 
parameters of the model, i.e., the sign of $b_c$ and the values 
of $(F,T,\beta)$. This understanding would provide the means for a comparison 
with the corresponding experimental results. Accordingly, we proceed 
by constructing ``state-diagrams'', which illustrate various domains 
in the parameter space (constrained by the range of experimental 
relevance, as discussed below). They correspond to each of the possible 
scenarios for the occurrence of sliding states.

In spite of the significant reduction in complexity provided by the 
representation as a linear superposition of solutions corresponding to three 
basic problems, it remains a challenge to study in detail a dependence on 
four parameters. Thus it is desirable to reduce the number of 
parameters, as well as the ranges within which the remaining ones are 
allowed to vary. For our system this reduction is feasible, based on the 
results of 
previously published studies, as well as based on the fact that some of the 
quantities involved in the definition of $F$ and $T$ are known material 
parameters. 

The available studies of Pt/PS or Pt/SiO$_2$ Janus colloids immersed in aqueous 
H$_2$O$_2$ solutions report the motion to be the one with the catalyst at 
the back (see, e.g., Refs. 
\citenum{Howse2007,Ebbens2013,Ebbens2019,Simmchen2016,Uspal2018b,Baraban2012,
Ilona2018}). The majority of these reports  
deal with particles presumably sliding along a horizontal wall, for which the 
velocity parallel to the wall is solely due to the chemical activity 
component (the sedimentation cannot contribute along that direction); Refs. 
\citenum{Ebbens2013,Ebbens2019,Brown2014} report also lift-off and upward 
motion, which can be due only to the chemical activity component. Accordingly, 
in the context of our phoretic motion models one infers that (see also Fig. 
\ref{fig3}) $\mathbf{U}_{ph}^{(fs)} \cdot \mathbf{p} < 0$. Based on the 
expression of $\mathbf{U}_{ph}^{(fs)}$ (Eq. (\ref{eq:Ufs})) the most 
plausible case\cite{Uspal2015a,Simmchen2016,Uspal2018b} is that the 
phoretic mobility 
of the active (catalyst-covered) area is negative, i.e., $b_c < 0$ 
(which corresponds to a repulsive effective interaction between the solute 
molecules and the particle) and $\mathcal{U}(\beta) > 0$. Therefore, in 
Eqs. 
(\ref{eq:scaled_Fg}) and (\ref{eq:scaled_Tg}) and 
(\ref{eq:dimless_motion_part})
\begin{equation}
\label{eq_signs_bc_Ub}
 b_c/|b_c| = -1\,,~\mathrm{and}~|\mathcal{U}(\beta)| = \mathcal{U}(\beta)\,,
\end{equation}
while from Eq. (\ref{eq:Ufs_models}) it follows that the parameter $\beta$ 
is restricted by the lower bound 
\begin{eqnarray}
 \label{eq:bound_beta}
 && -1 < \beta\,,~\textrm{constant flux,}\nonumber\\
 && -3.4 \lesssim \beta\,,~\textrm{variable flux.}
\end{eqnarray}
As far as an upper bound for the range of $\beta$ is concerned, for both 
models we shall explore values of $\beta$ as large as 
$\beta \simeq 1$, i.e., the same value of the phoretic mobility across the 
whole surface. For this value of $\beta$, the Janus particle does not 
exhibit phoretic rotations in response to distortions of the chemical field 
\cite{Uspal2015a,Popescu2018a}. Accordingly, this choice of the upper 
bound limits the study to the case that the rotation solely due to the 
distortion of the chemical field is such as to turn the 
active cap of the particle away from the wall \cite{Uspal2015a}. With reference 
to the left panel in Fig. \ref{fig1}, for a bottom heavy particle this 
situation promotes the emergence of sliding states -- upon approach to the 
wall -- with an orientation of the cap somewhat tilted away from the 
wall. This seems to be the situation in the experiments with Pt/PS and 
Pt/SiO$_2$ particles \cite{Simmchen2016,Baraban2012,Ebbens2019}.

Turning to the parameters $F$ and $T$, the mass densities of the core 
materials, of the catalyst, and of the solution, as well as the 
viscosity of the solution, are known. For the system of interest here, 
the particle parameters are $\rho_s \approx 1050$ kg/m$^3$ (PS) or 2196 kg/m$^3$ 
(amorphous SiO$_2$), and $\rho_a \approx 21450$ kg/m$^3$ (Pt catalyst), while 
for the aqueous solution $\rho \approx 1000$ kg/m$^3$ and $\mu \approx 9 \times 
10^{-4}$ Pa$\cdot$s (at 25 $^\circ$C). The radius of the particles is usually 
in the range $0.5~\mathrm{\mu m} \leq R \leq 3~\mathrm{\mu m}$. The parameter 
$\delta$ (i.e., the thickness of the catalyst layer at the pole of the 
Janus particle) is taken to be equal to the thickness reported by the thin-film 
deposition procedure, i.e., $\delta = \delta_{rep}$. Guided by the typical 
values reported in the literature 
\cite{Ilona2018,Baraban2012,Simmchen2016,Ebbens2013,Howse2007}, here we take 
$\delta$ in the range $7~\textrm{nm} \leq \delta \leq 30~\textrm{nm}$.

On the other hand, the dependence of $F$ and $T$ on the phoretic 
velocity $|\mathbf{U}_{ph}^{(fs)}|$ in unbounded solution is a point of 
concern, because this velocity is difficult to determine experimentally. Here, 
we simply assume, as done previously\cite{Simmchen2016,Uspal2018b,Uspal2018c}, 
that $|\mathbf{U}_{ph}^{(fs)}|$ is of the same order as the velocity observed in 
experiments for particles 
sliding along the wall. This interpretation is also supported by the 
results 
(reported in Ref. \citenum{Ebbens2013}) for the upward migration of Pt/PS 
particles (slightly density mismatched with the solution), which is similar to 
the values of the velocity along the wall. Furthermore, the 
theoretical predictions for the velocity $U_y$ along the wall (see Fig. 
\ref{fig3} and Eq. (\ref{eq:dimless_motion_part})) for the sliding 
states shown in Figs. \ref{fig5} - \ref{fig7}, as well as for a few other 
tested cases, have shown that it deviates by a factor $0.8 
- 1.1$ from the corresponding $|\mathbf{U}_{ph}^{(fs)}|$. Based on these 
arguments, we assume that the value of the phoretic velocity in unbounded 
solution lies in the range $1~\mathrm{\mu 
m/s} \leq |\mathbf{U}_{ph}^{(fs)}| \leq 10~\mathrm{\mu 
m/s}$, as usually reported by experimental studies. \bibnote{Here we 
note that much larger velocities, of the order of tens (up to a hundred) 
of $\mathrm{\mu m/s}$, have been reported in Ref. \citenum{Isa2017} for Pt/PS 
spherical Janus particles in aqueous peroxide solutions of moderate 
concentrations. These values are very different from all the other 
reports in the literature (for seemingly similar 
systems)\cite{Baraban2012,Howse2007,Howse2015,Brown2014,Ebbens2013,Ebbens2019}. 
Accordingly, we leave aside this exceptional, rather than typical case.}

%%%%%%%%%%%%%%%%%%%%%%%%%%%%%%%%%%%%%%
\begin{figure}[!b]
    \centering
    \includegraphics[width= \columnwidth]{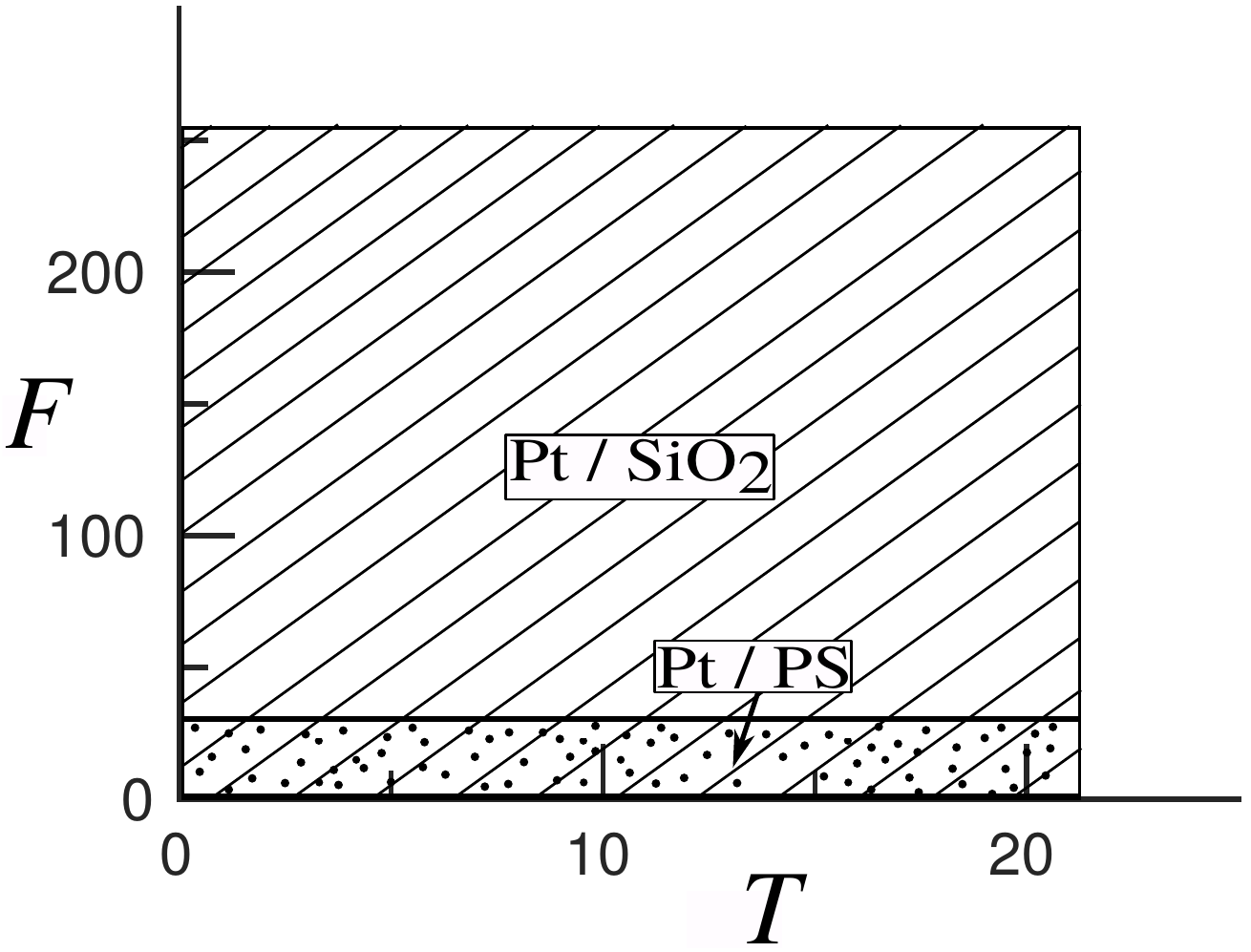}
     \caption{
\label{fig:param_ranges}
{\small 
The subspace of the parameter space $(F,T)$ (see Eq. (\ref{eq:defs_F_T})) 
which is relevant for experiments with Pt/PS and Pt/SiO$_2$ chemically 
active Janus particles immersed in aqueous H$_2$O$_2$ solutions (see the main 
text).
}
}
\end{figure}
%%%%%%%%%%%%%%%%%%%%%%%%%%%%%%%%%%%%%%
Collecting all pieces, one arrives at the experimentally relevant 
ranges for the parameters $F$ and $T$ as
\begin{eqnarray}
\label{eq:F_T_ranges}
 && 0.11 \lesssim F \lesssim 254 \label{eq:F_range}\\
 && \hspace*{-3.2cm} \mathrm{and}\nonumber\\
 && 0.09 \lesssim T \lesssim 21 \label{eq:T_range}\,.
\end{eqnarray}
This parameter subspace is shown in Fig. \ref{fig:param_ranges}, where the 
relatively small subset corresponding to the particles with PS core (for 
which there are more experimental reports available) is also indicated.

%%%%%%%%%%%%%%%%%%%%%%%%%%%%%%%%%%%%%%
\begin{figure*}[!h]
    \centering
    \includegraphics[width=.9\textwidth]{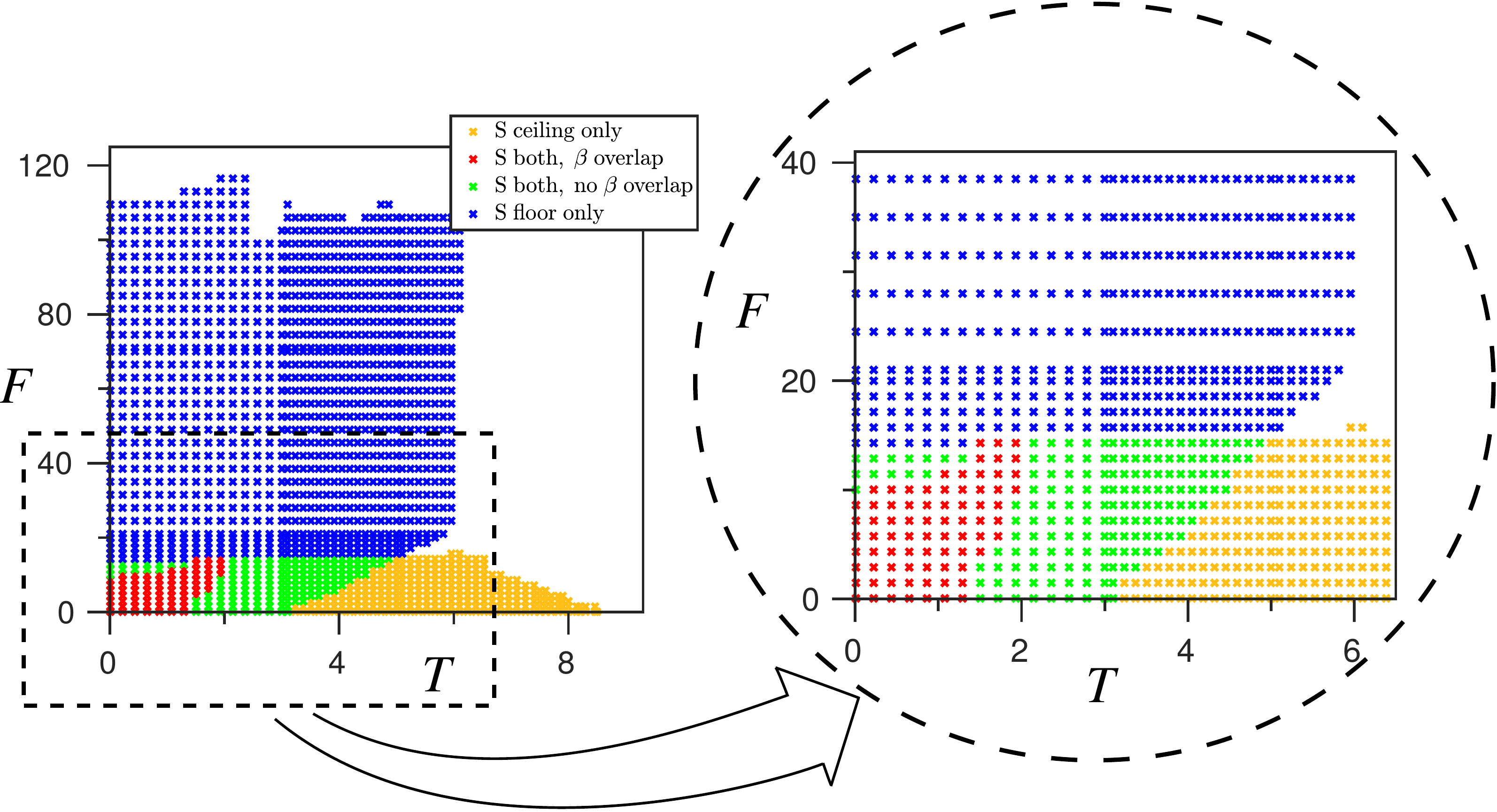}
     \caption{
\label{fig_state_dia_cf}
{\small 
State-diagram of sliding states (S) for the constant flux model and $b_c < 0$. 
The white region corresponds to the absence of sliding states at either 
wall, within the range of values of $\beta$ analyzed here (see 
the main text). The label ``$\beta$ overlap'' and ``no $\beta$ overlap''
refers to scenario (v) and (iv), respectively, described in the main text.
}
}
\end{figure*}
%%%%%%%%%%%%%%%%%%%%%%%%%%%%%%%%%%%%%%
The study of the dynamics as a function of the parameters $F$, $T$, and $\beta$, 
constrained to the ranges outlined above, has been carried out for each of 
the two model chemical activities as follows. A grid of pairs $(F,T)$, spanning 
the whole relevant domain (Eqs. (\ref{eq:F_range}) and (\ref{eq:T_range})), is 
selected; for each pair $(F,T)$, phase portraits as the ones above are 
generated -- for both the ceiling and the floor cases -- at 
a number of values of $\beta$, in steps of 0.1 starting from the minimum within 
the corresponding range (Eq. (\ref{eq:bound_beta})). These phase portraits are 
analyzed with 
regard to the occurrence of sliding states; the outcome is used to 
classify the behavior at $(F,T)$ as follows (see also 
SI Sec. \ref{sec:appA}): (i) no sliding states occur, irrespective of 
$\beta$ and the 
kind of wall (i.e., floor or ceiling); (ii) sliding states 
occur for a subset of values of $\beta$, but \textit{only} at the 
floor; (iii) sliding states occur, for a subset of values of  
$\beta$, but \textit{only} at the ceiling; (iv) sliding states occur 
\textit{both} at the floor and at the ceiling, but the subsets of values 
of $\beta$ for the floor and for the ceiling do not overlap; (v) sliding 
states occur \textit{both} at the floor and at 
the ceiling, and the intersection of the subsets of values of $\beta$ 
for the floor and for the ceiling is non-void. (This listing includes all 
scenarios encountered in the theoretical analysis.) 

\begin{figure*}[!htb]
    \centering
    \includegraphics[width=.9\textwidth]{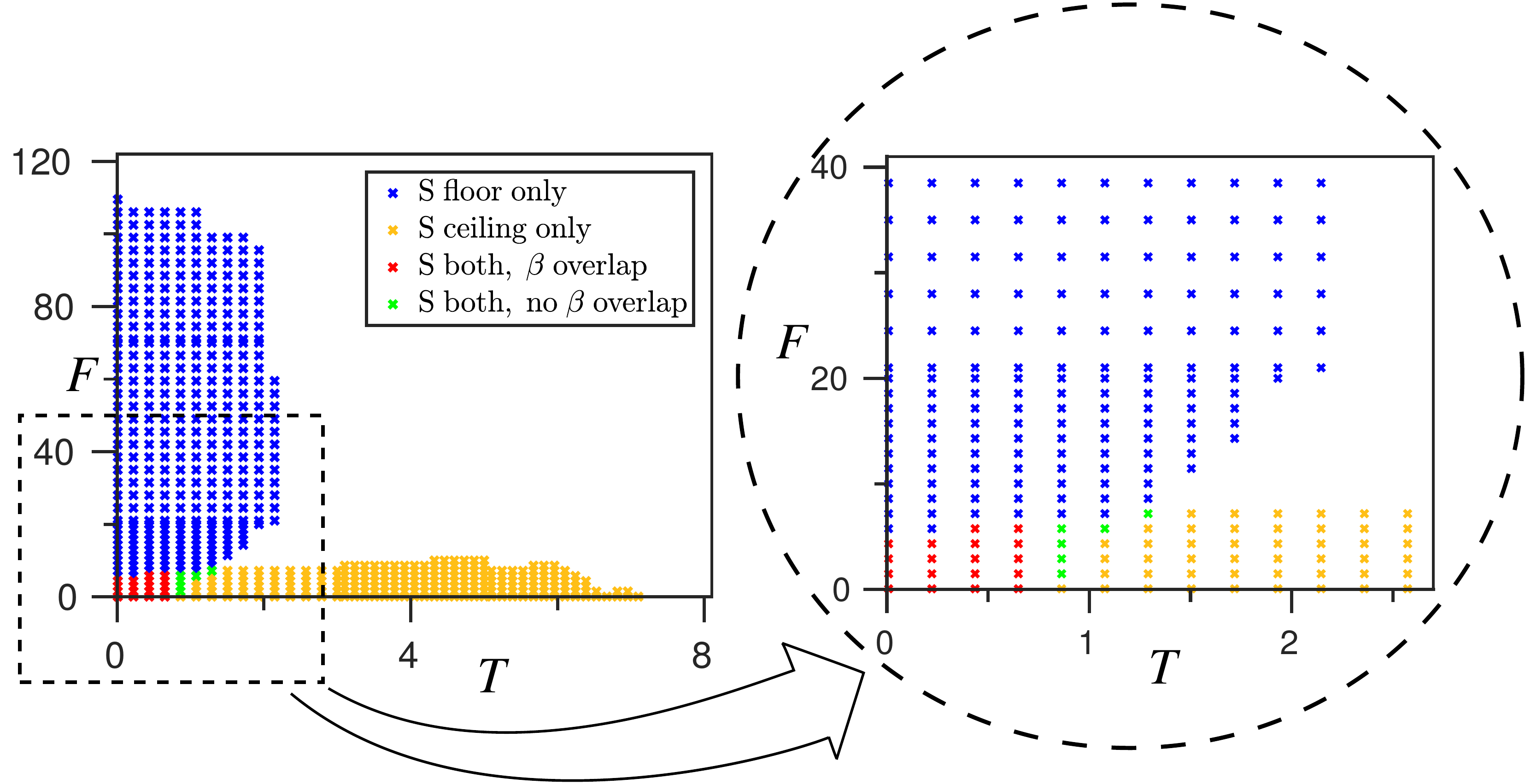}
     \caption{
\label{fig_state_dia_vf}
{\small 
State-diagram of sliding states (S) for the variable flux model and $b_c < 0$. 
The white region corresponds to the absence of sliding states at either 
wall, within the range of values of $\beta$ analyzed here (see 
the main text). The label ``$\beta$ overlap'' and ``no $\beta$ overlap''
refers to scenario (v) and (iv), respectively, described in the main text.
}
}
\end{figure*}

The resulting classes, i.e., $(F,T)$ and one of the five cases above, provide a 
``state-diagram'' of the sliding steady-states of the dynamics. The 
state-diagrams for the constant-flux (``cf'') and the variable-flux 
(``vf'') models, which are the main results of the present study, are shown in 
Figs. \ref{fig_state_dia_cf}  and  \ref{fig_state_dia_vf}, respectively.
%%%%%%%%%%%%%%%%%%%%%%%%%%%%%%%%%%%%%%
%%%%%%%%%%%%%%%%%%%%%%%%%%%%%%%%%%%%%%
%%%%%%%%%%%%%%%%%%%%%%%%%%%%%%%%%%%%%%
\begin{figure*}[!t]
    \centering
    \includegraphics[width=.95\textwidth]{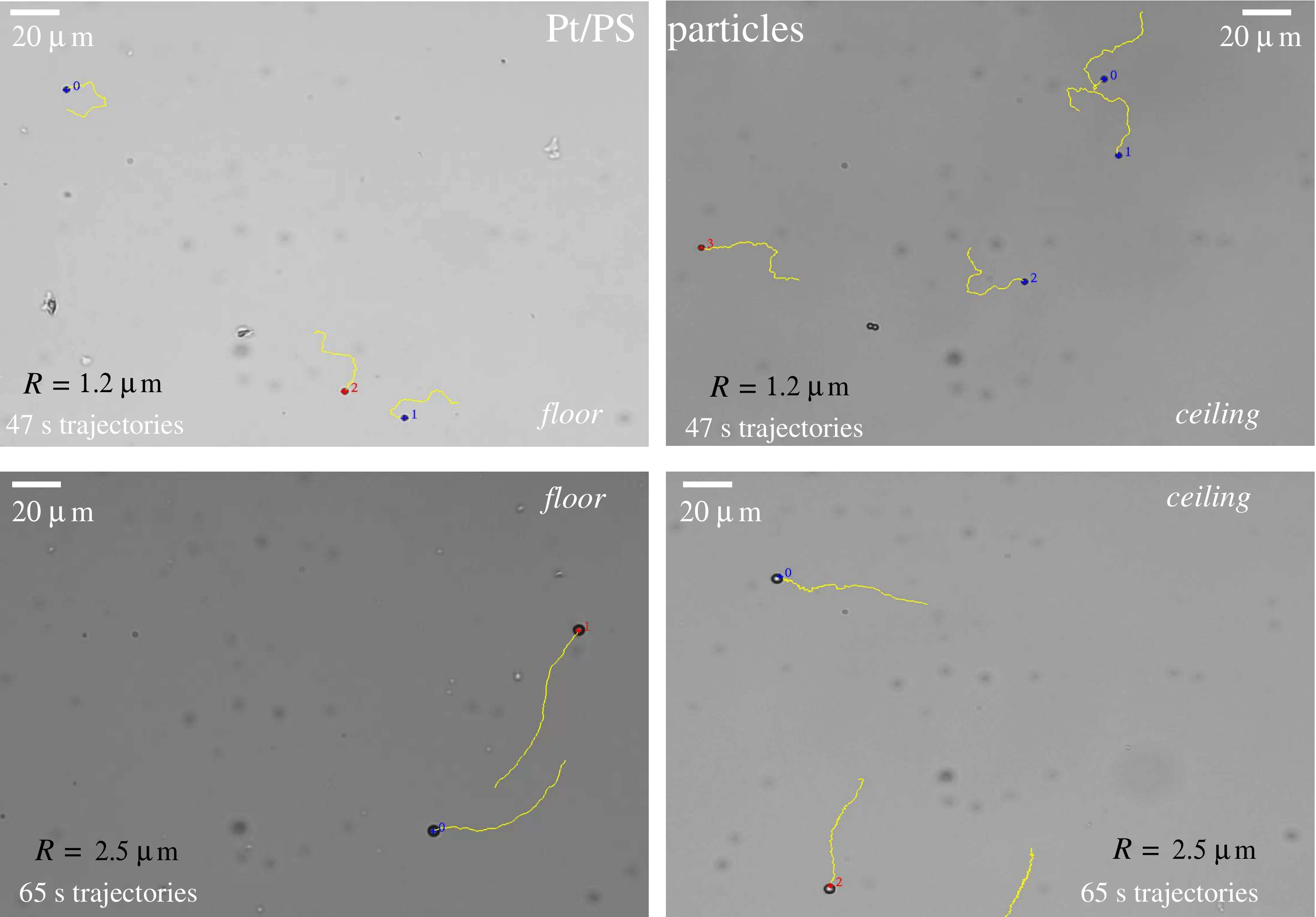}
     \caption{
\label{fig:tracked_traj_PS}
{\small 
Tracked trajectories of Pt/PS Janus particles in sliding states at the floor 
(left column) and at the ceiling (right column), for particles of size $R = 
1.2~\mathrm{\mu m}$ (top row) and $2.5~\mathrm{\mu m}$ (bottom row), 
respectively. The progression of the trajectory is from its free end 
towards the end with the particle attached. Note the difference in 
duration of the trajectory between the top and bottom rows. The labels near 
the particles are added by the tracking software; their sole purpose is that of 
indexing the various tracked trajectories.
}
}
\end{figure*}
%%%%%%%%%%%%%%%%%%%%%%%%%%%%%%%%%%%%%%
At the quasi-quantitative level, these state-diagrams reveal significant 
differences between the behaviors emerging from the two models. Although 
qualitatively they cannot be discriminated, because both capture all 
possible outcomes (i)-(v), the domains corresponding to these states differ 
significantly; in particular, compared with model ``cf'', model ``vf'' 
predicts a significantly narrowed range of values of $T$ for which 
sliding states at the floor may occur (the blue region), as well as a 
significant shrinking of the domain where coexisting sliding states may occur 
(the red region).

\subsection{\large Comparison with experimental results}
By comparing Figs. \ref{fig_state_dia_cf} and \ref{fig_state_dia_vf} with 
Fig. \ref{fig:param_ranges}, one infers that Pt/PS Janus particles are the 
best candidates for the experimental validation of coexisting sliding 
states because their parameter $F$ is typically within the range 
where such state are expected to occur. This is a somewhat fortunate 
situation, in view of the fact that a number of experimental studies employing 
Pt/PS Janus particles are available 
\cite{Baraban2012,Brown2014,Howse2015,Ebbens2019} and, as noted 
above, in some of them the presence of motile particles at the top 
walls was noted \cite{Brown2014,Howse2015}. On the other hand, for Pt/SiO$_2$ 
Janus particles the mass density mismatch between the core material and the 
solution is large, and thus $F$ attains larger values. Accordingly, such 
particles may be used to explore the region of coexisting sliding 
states only if they are small and if their self-propulsion is sufficiently 
strong. In practice, the latter condition might be difficult to 
achieve because it requires a relatively large rate of solute (O$_2$) 
production and the formation of O$_2$ bubbles in the solution becomes 
significant. Guided by these observations, our experiments have been 
focused on the case of Pt/PS Janus particles; but, as discussed in the 
experimental section, Pt/SiO$_2$ Janus particles of two different sizes have 
been studied, too. Further motivation for carrying out these experiments 
follows from the fact that the number of independent studies of such particles 
is quite small yet, and that there are large discrepancies in the 
values of the velocity reported for presumably similar experiments performed by 
different groups. For example, for Pt/PS particles of radius 1 $\mathrm{\mu m}$ 
in 10\% H$_2$O$_2$ aqueous solution, Ref. \citenum{Brown2014} reports a velocity 
of 15 - 20 $\mathrm{\mu m/s}$, and Ref. \citenum{Howse2015} quotes a 
velocity slightly less than 4 $\mathrm{\mu m/s}$, while the only 
differences between these particles consists of the thickness of the Pt 
layer (5 nm vs 10 nm) and of the method of depositing Pt. This 
difference by a factor of 4-5, which directly transfers into the magnitudes 
of $F$ and $T$, can cause significant differences concerning the region 
in the state diagram which the point should be attributed to (see also, c.f., 
Fig. \ref{fig:exp_vs_theo}).

Both the smaller ($R = 1.2~\mathrm{\mu m}$) and the larger ($R = 
2.5~\mathrm{\mu m}$) Pt/PS Janus particles, which self-propel within 
the $3 \%$ H$_2$O$_2$ aqueous solution, exhibit sliding states at both the 
ceiling and the floor (see the videos SI.V1 to SI.V4), so that these 
states coexist. This finding provides a welcome reassurance that the 
phenomenon is indeed experimentally observable and robust with respect 
to variations of the parameters. As illustrated by the tracked 
trajectories shown in Fig. \ref{fig:tracked_traj_PS} (see also the videos SI.V1 
to SI.V4), the sliding motions at the ceiling and at the floor are very 
similar. (We note that the recordings cover different time 
intervals. Accordingly, the duration of the trajectories in the top panels 
in Fig. \ref{fig:tracked_traj_PS} is different from that of the trajectories in 
the bottom panels in Fig. \ref{fig:tracked_traj_PS}). Therefore, as noted 
in the experimental section, the velocities of the particles for the sliding 
states are de facto independent of the location (ceiling or 
floor) and of the size of the particles. Furthermore, in the videos SI.V4 and 
SI.V5 (see the SI) one clearly observes that particles arrive from 
the bulk and start to slide along the ceiling (see the particles 
labeled as ``2'' in each of these videos, and the description of the videos 
in the SI Sec. \ref{sec:appD}). This observation confirms that the 
presence of the particles in 
sliding states at the ceiling is the result of the dynamics and not some 
artifact such as the attachment of the particles when the ceiling (i.e., a 
glass slide) is placed at the top of the experimental cell. Both the observed 
phenomenon and the value of the velocity (comparable at the floor 
and at the ceiling) of the particles are commensurate with 
those noted in Ref. \citenum{Howse2015} for a similar concentration of 
the aqueous H$_2$O$_2$ solution.
%%%%%%%%%%%%%%%%%%%%%%%%%%%%%%%%%%%%%%
\begin{figure}[!th]
    \centering
    \includegraphics[width=.95\columnwidth]{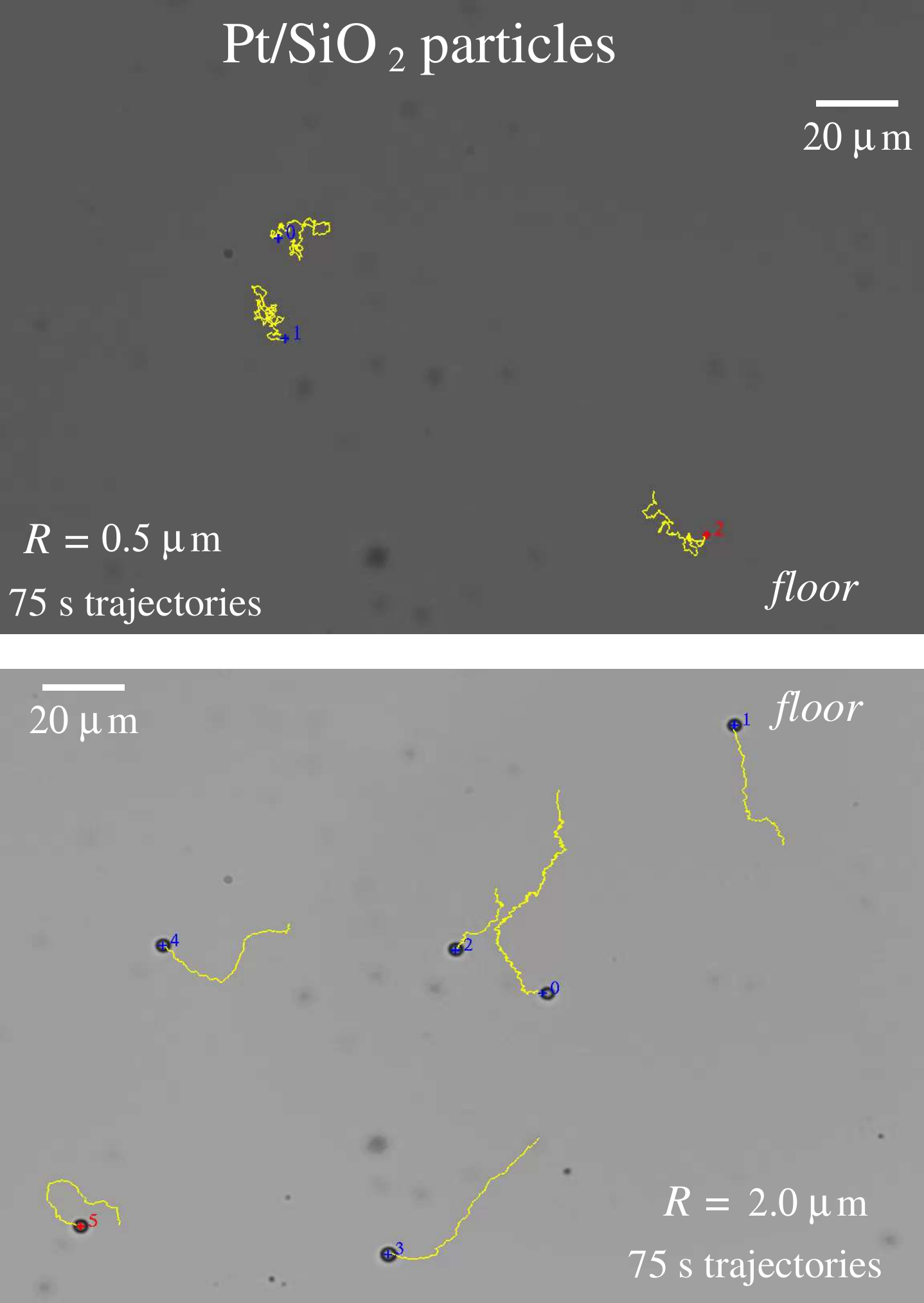}
     \caption{
\label{fig:tracked_traj_Si}
{\small 
Tracked trajectories of Pt/SiO$_2$ in sliding states at the floor, 
for particles of sizes $R = 0.5~\mathrm{\mu m}$ (top panel) and 
$2.0~\mathrm{\mu m}$ (bottom panel), respectively. The progression of 
the trajectory is from its free end towards the end with the particle 
attached. There is a significant winding of the trajectories of the 
smaller particles compared to the ones of the larger particles. 
The labels near the particles are added by the tracking software; their sole 
purpose is that of indexing the various tracked trajectories.
}
}
\end{figure}
%%%%%%%%%%%%%%%%%%%%%%%%%%%%%%%%%%%%%%

%%%%%%%%%%%%%%%%%%%%%%%%%%%%%%%%%%%%%%%%%%%%%%%%%%%%%%%%%%
\begin{table*}[!t]
\scalebox{0.9}{
\begin{tabular}{|c|c|c|c|c|c|c|c|c|c|c|}
\hline
\multirow{2}{*}{\texttt{source}} & 
\multirow{2}{*}{\texttt{\begin{tabular}[c]{@{}c@{}}particle\\ 
type\end{tabular}}} & 
\multirow{2}{*}{\textbf{\begin{tabular}[c]{@{}c@{}}$R$\\ 
($\mu m$)\end{tabular}}} & 
\multirow{2}{*}{\textbf{\begin{tabular}[c]{@{}c@{}}$\delta$\\ 
($nm$)\end{tabular}}} & 
\multirow{2}{*}{\texttt{\begin{tabular}[c]{@{}c@{}}sliding\\ 
location\end{tabular}}} &
\multirow{2}{*}{\textbf{\begin{tabular}[c]{@{}c@{}}$U_{ph}^{(fs)}$\\ ($\mu 
m/s$)\end{tabular}}} & 
\multirow{2}{*}{\textit{\textbf{$F$}}} & 
\multirow{2}{*}{\textit{\textbf{$T$}}} & 
\multirow{2}{*}{\texttt{symbol}}       
                                          & 
\multicolumn{2}{c|}{\texttt{\begin{tabular}[c]{@{}c@{}} location 
check \end{tabular}}} \\ \cline{10-11} 
                                  &                                             
 
                                     &                                          
 
                                      &                                         
 
                                           &                                    
&                                                                               
 
                 &                                      &                       
 
              &                                                                 
 
                      & \texttt{model $cf$}                      & 
\texttt{model $vf$}                      \\ \hline
this work                         & $\rm{Pt/PS}$

                                     & 1.2                                      
 
                                      & 9                                       
 
                                           & $(f,c)$                            
& 1.6                                                                           
 
                 & 2.5                                  & 1.6                   
 
              & $\color{red}\tikzsymbol[rectangle]{minimum 
width=6pt,minimum height=6pt,fill=yellow}$  & $\checkmark$                     
             & $\times$                                     \\ \hline
this work                         & $\rm{Pt/PS}$

                                     & 2.5                                      
 
                                      & 9                                       
 
                                           & $(f,c)$                            
& 1.6                                                                           
 
                 & 7.4                                  & 3.3                   
 
              & $\color{blue}\tikzsymbol[rectangle]{minimum width=6pt,minimum 
height=6pt,fill=yellow}$ & $\times$                                      & 
$\times$                                     \\ \hline
this work                         & $\rm{Pt/SiO_2}$

                                     & 0.5                                      
 
                                      & 9                                       
 
                                           & $(f)$                              
& 2.1                                                                           
 
                 & 3.6                                  & 0.5                   
 
              & $\color{blue}\tikzcircle[fill=green]{3.5pt}$                    
                         & $\times$                                      & 
$\times$                                     \\ \hline
this work                         & $\rm{Pt/SiO_2}$

                                     & 2                                        
 
                                      & 9                                       
 
                                           & $(f)$                              
& 1.4                                                                           
 
                 & 77.3                                 & 3                     
 
              & $\color{red}\tikzcircle[fill=green]{3.5pt}$                     
                         & $\checkmark$                                  & 
$\times$                                     \\ \hline
Ref. \citenum{Simmchen2016}                              & $\rm{Pt/SiO_2}$      
 
                                                            & 1                 
 
                                                             & 7                
 
                                                                  & $(f)$       
 
                      & 6                                                       
 
                                       & 4.6                                  & 
0.27                                 & $\bf{\medtriangleleft}$                  
 
                                             & $\times$                         
 
            & $\times$                                     \\ \hline
Ref. \citenum{Simmchen2016}                              & $\rm{Pt/SiO_2}$      
 
                                                            & 2.5               
 
                                                             & 7                
 
                                                                  & $(f)$       
 
                      & 6                                                       
 
                                       & 27.8                                 & 
0.7                                  & $\medtriangleleft$                       
 
                                             & $\checkmark$                     
 
            & $\checkmark$                                 \\ \hline
Ref. \citenum{Brown2014}                              & $\rm{Pt/PS}$            
 
                                                         & 1                    
 
                                                          & 5                   
 
                                                               & $(f,c)$        
 
                   & 17                                                         
 
                                    & 0.13                                 & 
0.07                                 & $\medcircle$                             
 
                                             & $\checkmark$                     
 
            & $\checkmark$                                 \\ \hline
Ref. \citenum{Baraban2012}                               & $\rm{Pt/PS}$         
 
                                                            & 2.5               
 
                                                             & 7                
 
                                                                  & $(f)$       
 
                      & 2.5                                                     
 
                                       & 4.3                                  & 
1.6                                  & $\medtriangleright$                      
 
                                             & $\checkmark$                     
 
            & $\times$                                     \\ \hline
Ref. \citenum{Baraban2012}                               & $\rm{Pt/PS}$         
 
                                                            & 2.5               
 
                                                             & 7                
 
                                                                  & $(f)$       
 
                      & 9                                                       
 
                                       & 1.2                                  & 
0.45                                 & $\medtriangleright$                      
 
                                             & $\times$                         
 
            & $\times$                                     \\ \hline
Ref. \citenum{Howse2015}                              & $\rm{Pt/PS}$            
 
                                                         & 1                    
 
                                                          & 7                   
 
                                                               & $(f,c)$        
 
                   & 1                                                          
 
                                    & 2.6                                  & 
1.6 
                                 & $\meddiamond$                                
 
                                         & $\checkmark$                         
 
        & $\times$                                     \\ \hline
Ref. \citenum{Howse2015}                              & $\rm{Pt/PS}$            
 
                                                         & 1                    
 
                                                          & 7                   
 
                                                               & $(f,c)$        
 
                   & 3.75                                                       
 
                                    & 0.7                                  & 
0.4 
                                 & $\meddiamond$                                
 
                                         & $\checkmark$                         
 
        & $\checkmark$                                 \\ \hline
Ref. \citenum{Howse2015}                              & $\rm{Pt/PS}$            
 
                                                         & 1.5                  
 
                                                          & 7                   
 
                                                               & $(f,c)$        
 
                   & 1                                                          
 
                                    & 4.8                                  & 
2.5 
                                 & $\meddiamond$                                
 
                                         & $\times$                             
 
        & $\times$                                     \\ \hline
Ref. \citenum{Howse2015}                              & $\rm{Pt/PS}$            
 
                                                         & 1.5                  
 
                                                          & 7                   
 
                                                               & $(f,c)$        
 
                   & 2.5                                                        
 
                                    & 1.9                                  & 
1.0 
                                 & $\meddiamond$                                
 
                                         & $\checkmark$                         
 
        & $\checkmark$                                 \\ \hline
Ref. \citenum{Howse2015}                              & $\rm{Pt/PS}$            
 
                                                         & 2.5                  
 
                                                          & 7                   
 
                                                               & $(f,c)$        
 
                   & 1                                                          
 
                                    & 10.7                                 & 
4.1 
                                 & $\meddiamond$                                
 
                                         & $\times$                             
 
        & $\times$                                     \\ \hline
Ref. \citenum{Ebbens2019}                              & $\rm{Pt/PS}$           
 
                                                          & 3.5                 
 
                                                           & 10                 
 
                                                                & $(c)$         
 
                    & 5.4                                                       
 
                                     & 3.9                                  & 
1.5                                  & $\medsquare$                             
 
                                             & $\checkmark$                     
 
            & $\checkmark$                                 \\ \hline
\end{tabular}
\caption{\label{table:1} The values of the parameters corresponding to 
the experimentally studied particles indicated by symbols in Fig.  
\ref{fig:exp_vs_theo}. The letters \textit{f} and \textit{c} in the 
fifth column refer to sliding states located near the floor or the ceiling, 
respectively. The tick ``$\checkmark$'' indicates that the 
corresponding point belongs to a compatible domain of the 
theoretically predicted state-diagram for the corresponding model (i.e., 
$cf$ (constant flux) or $vf$ (variable flux), 
respectively), whereas the symbol ``$\times$'' indicates that 
it does not belong to them.
}
}
\end{table*}
%%%%%%%%%%%%%%%%%%%%%%%%%%%%%%%%%%%%%%%%%%%%%%%%%%%
In contrast, at the same concentration of the H$_2$O$_2$ aqueous solution, in 
the case of the Pt/SiO$_2$ Janus particles we have observed sliding states only 
at the floor, even when using smaller particles ($R = 0.5~\mathrm{\mu m}$). 
The typical motion in a sliding state at the floor is illustrated, in the 
form of tracked trajectories, in Fig. \ref{fig:tracked_traj_Si} (see also the 
videos SI.V6 and SI.V7 in the SI). The difference in speed between the 
particles of different sizes (see the experimental section), with the 
smaller particles moving faster (see also Ref. \citenum{Golestanian2012}), 
is not easily noticeable via visual inspection. 

Finally, we note that while, as discussed above, the models capture all 
the relevant, experimentally observable phenomenology, it is interesting to 
analyze whether these considerations hold even on a more quantitative 
level. For example, a simple check is provided by locating the 
points $(T,F)$ corresponding to the experiments (the present ones and 
from, e.g., Refs. \citenum{Baraban2012,Brown2014,Howse2015,Ebbens2019}) within 
the state-diagrams and by checking the compatibility between the observed 
state(s) and the model-predicted state. This is illustrated in Fig. 
\ref{fig:exp_vs_theo}, where we show the points, corresponding to our 
estimates $(T,F)$ for the various experimental realizations, on top of the 
region of the state-diagram to which they belong to (see Table \ref{table:1}).

%%%%%%%%%%%%%%%%%%%%%%%%%%%%%%%%%%%%%%
\begin{figure*}[!t]
    \centering
    \includegraphics[width=.99\textwidth]{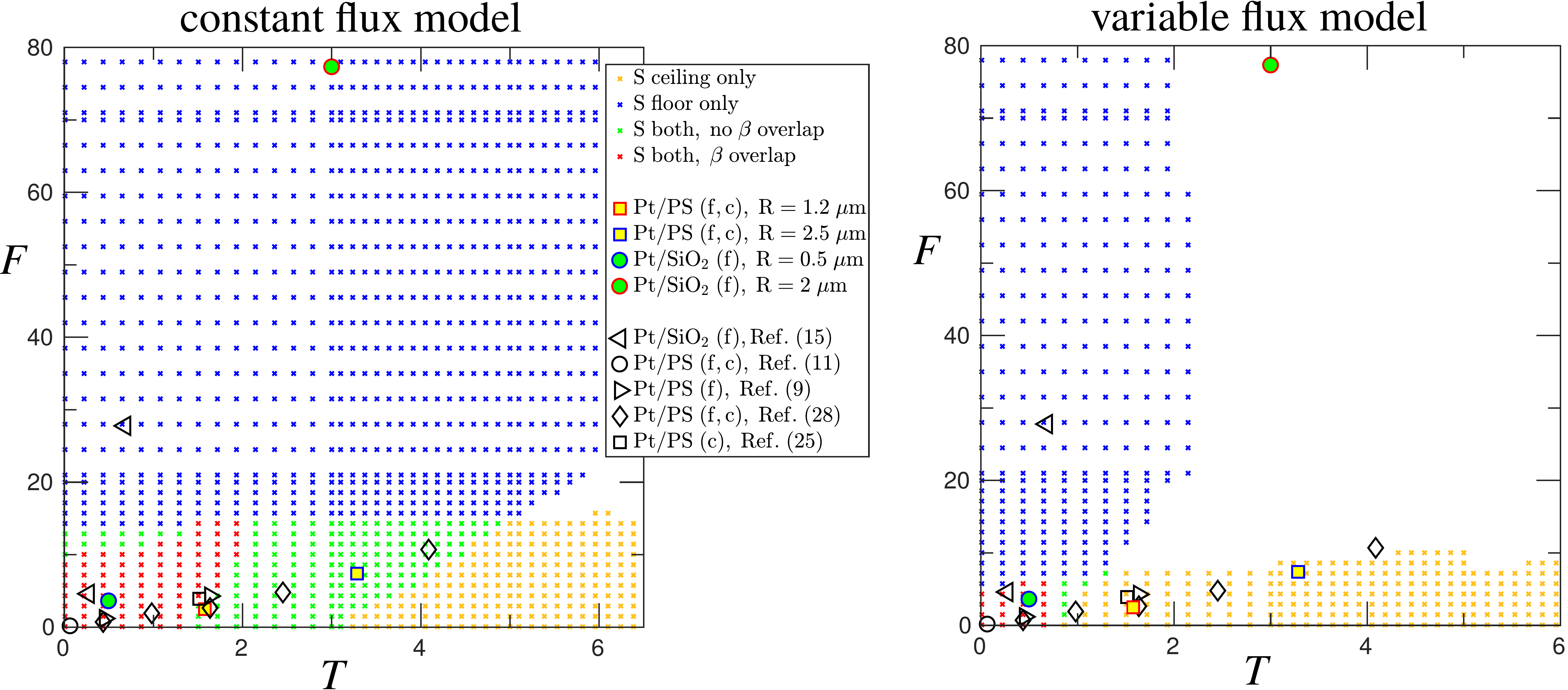}
     \caption{
\label{fig:exp_vs_theo}
{\small 
Comparison of the experimental results (symbols, see the legend valid for 
both panels) with the corresponding theoretical predictions of their state (in 
terms of sliding state occurrence). The indices $c$ and $f$ refer to the 
location (ceiling or floor) where the experimental observation was made. The 
theoretical predictions shown in the two panels are the corresponding 
sub-regions indicated in Figs. \ref{fig_state_dia_cf} and 
\ref{fig_state_dia_vf}. The label ``$\beta$ overlap'' and ``no $\beta$ 
overlap'' refers to scenario (v) and (iv), respectively, described in the 
main text.
}
}
\end{figure*}
%%%%%%%%%%%%%%%%%%%%%%%%%%%%%%%%%%%%%%   
It is apparent that, at the quantitative level, neither of the 
two models captures the complete set of experimental data points as a 
whole. However -- recalling the simplicity of the models employed in 
the study -- the agreement is actually reasonably good. For example, 
concerning the constant flux model the predictions are in agreement with the 
experimental observations in nine out of the fifteen cases studied here 
(see Table \ref{table:1}). (It has to be noticed that in three of these 
cases observations have been reported either only at the floor or only at the 
ceiling. Accordingly, only a somewhat weak agreement can be claimed because, in 
the absence of observations at both locations, it cannot be inferred whether 
the experimental point belongs to one of the regions without 
coexistence (blue, yellow, or green), or to the one with coexistence 
(red).) Furthermore, from the point of view of these state-diagrams, it may 
eventually be inferred that the ``cf'' model seems to provide a closer match 
of the experimental observations than the ``vf'' model does. However, 
before 
drawing 
conclusions such as that one or the other (or both) activity models 
are too simple (i.e., they miss essential physical 
ingredients), one should recall that very crude models and 
approximations have been made concerning 
the estimates of the parameters $\delta$ and $|\mathbf{U}_{ph}^{(fs)}|$. (In 
addition, the latter is further affected by conflicting values reported from 
various experiments.) Taking the case of Pt/PS particles and 
the 
coexisting sliding states as an example, one observes that increasing the 
estimates of $\delta$ by a factor of, e.g., 1.5 is sufficient to shift one of 
the points corresponding to the current experiments (filled squares) and some 
of those in Ref. \citenum{Howse2015} (open diamonds) out of the 
red colored region of the state-diagram corresponding to the constant 
flux model. This sensitivity concerning the precise value of $\delta$ is 
heightened further by noting that its interpretation is connected with a very 
simple geometrical model for the catalyst film, the reliability of which has 
only recently been been investigated experimentally \cite{Wirth2018}. 
Accordingly, while the results of the present analysis 
are encouraging, we think that they are insufficient for drawing strong 
conclusions, such as claims of quantitative agreement or 
discriminating resolution between distinct models of chemical 
activity.  
 
\section{Conclusions}

We have theoretically investigated the dynamics of chemically active, 
gyrotactic Janus particles near walls. The particles are located either 
above the floor or below the ceiling of the sample. We have focused on 
the emergence of wall-bound steady sliding states. For this analysis, we 
have used two distinct models corresponding to a constant (cf) 
or a variable (vf) flux of the  chemical emission of the particle. 
The theoretical analysis has been complemented by experiments conducted with 
Pt/PS and Pt/SiO$_2$ Janus particles immersed in aqueous H$_2$O$_2$ solutions, 
which have 
been set up such as to optimize the eventual occurrence of sliding states 
simultaneously at the floor and the ceiling. We have shown that, for both 
choices of the chemical activity, the models capture all the phenomenology of 
sliding states observed in the experiments, including that of coexistence of 
ceiling- and floor-sliding states as noted in previous experimental studies 
and confirmed by the present Pt/PS experiments.

For each of the two models, the various scenarios of sliding states at 
the ceiling and the floor have been studied as functions of the 
parameters $(F,T,\beta)$ that determine the dynamics (see Eq. 
(\ref{eq:dimless_motion_part})). The results concerning the sliding 
state occurrence have been summarized in terms of ``state-diagrams'' 
 in the $(T,F)$ plane. In particular, the occurrence of sliding state 
coexistence at the ceiling and the floor, or the occurrence of only 
one of the two states, at points $(T,F)$ where both states are allowed 
but for 
different values of $\beta$, sets very strong bounds on the otherwise hard 
to measure (or estimate) ratio $\beta$ of the phoretic 
mobilities for self-phoretic Janus particles (Eq. (\ref{eq:def_beta})).

The structure of these ``state-diagrams'' seems to be sufficiently 
different to allow one, eventually, to discriminate between models 
with different choices of the chemical activity (or ruling out both of them), 
provided quantitative comparisons can 
be made with the experiments. A first attempt of such a comparison 
with the 
results of present experiments, as well as with those of previously 
published 
experimental studies, shows a reasonable quasi-quantitative 
agreement with the theoretical predictions. However, we caution that this 
finding could be 
a spurious feature emerging from the inherent uncertainties in estimating the 
phoretic velocity in unbounded solutions and in modeling the distribution of 
catalyst at the surface of the particles. In particular for Pt/PS particles, 
the relevant parameters characterizing them depend very sensitively on these 
two quantities. This uncertainty, together with the 
significant differences between the velocities reported in the available 
experimental investigations for seemingly similar particles and solutions, 
highlights the necessity of additional, systematic, and thorough 
experimental 
studies. The state-diagrams derived here, and the underlying 
analysis of 
the role of the various parameters and of the assumptions concerning 
the geometry of the problem, should prove to be useful in guiding such 
investigations. The results of such 
studies may then eventually validate and discriminate between the various models 
of the self-motility mechanism.

\begin{suppinfo}
Video recordings, showing the motion, within a 3\% (v/v) aqueous H$_2$O$_2$ 
solution, of Pt/PS particles at the floor and at the ceiling, and of 
Pt/SiO$_2$ particles at the floor, are provided as Supplementary 
Information (SI). Additionally, the SI contains details concerning the 
procedure to classify the ``state scenario'' at a point $(F,T)$ (SI Sec. 
\ref{sec:appA}), a discussion of the observation of ``hovering'' steady-states 
at the floor and at the ceiling (SI Sec. \ref{sec:appB}), a 
glossary and list of symbols (SI Sec. \ref{sec:appC}), and a brief description 
of the video files (SI Sec. \ref{sec:appD}).
\end{suppinfo}

\bigskip

\noindent\textbf{~~Notes}\newline
The authors declare that there is no competing financial interest.\newline


\begin{acknowledgement}
I.K. and Z.J. acknowledge support by the National Science Foundation under 
the award number CBET-1705565.
\end{acknowledgement}

% \begin{suppinfo}
% 
% A listing of the contents of each file supplied as Supporting Information
% should be included. For instructions on what should be included in the
% Supporting Information as well as how to prepare this material for
% publications, refer to the journal's Instructions for Authors.
% 
% The following files are available free of charge.
% \begin{itemize}
%   \item Filename: brief description
%   \item Filename: brief description
% \end{itemize}
% 
% \end{suppinfo}


\clearpage

\bibliography{refs_top_bottom_sliding_v4}

\clearpage

\appendix

\counterwithin{figure}{section}

%\onecolumn

\twocolumn[
  \begin{@twocolumnfalse}
 \centerline{\Large \textbf{Supplementary Information}}
  \end{@twocolumnfalse}
\vspace*{0.5 in}  
]

\section{\label{sec:appA}Classifying the ``state 
scenario'' at a point $(F,T)$}

\begin{figure}[!b]
	\centering
	\includegraphics[width=1.\columnwidth]{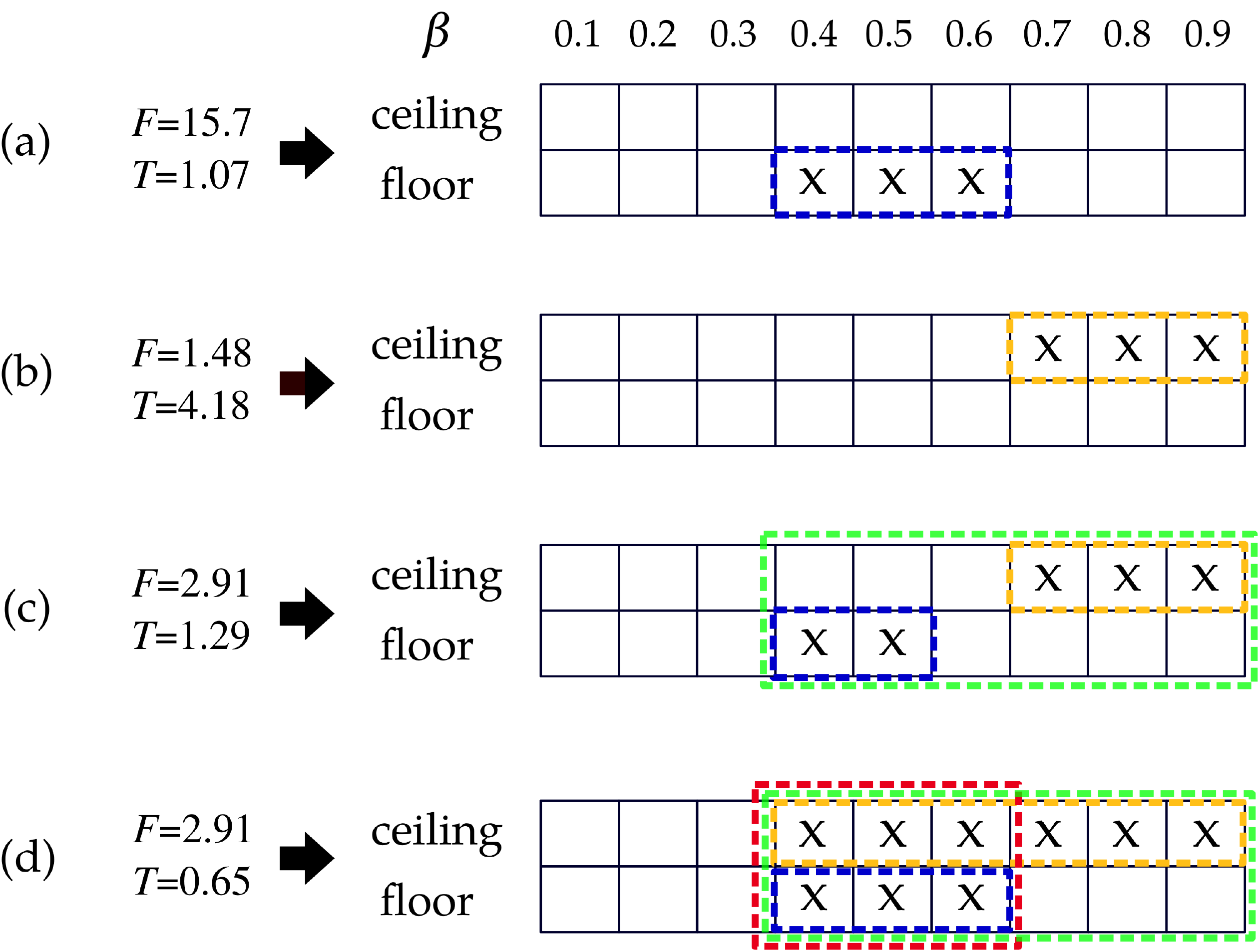}
	\caption{\label{fig:App_A}
	{\small 
	Schematic illustration of the interpretation of the ``states'' in 
terms of sliding states at the floor and the ceiling for various 
values of $\beta$ (here, for reasons of clarity, only the range $0.1 
\leq \beta \leq 0.9$ is shown). The various panels correspond to the specific 
points 
$(F,T) = $ (15.7,1.07) (a); (1.48,4.18) (b); (2.91,1.29) (c); and (2.91,0.65) 
(d), which are located in the four distinct regions in Fig. 
\ref{fig_state_dia_cf}.
}
}
\end{figure}

As explained in the main text, at a given point $(F,T)$ in the parameter 
space, $\beta$ is varied within the corresponding range of 
values, and the phase portraits are analyzed with respect to the 
occurrence of sliding states. These results are algorithmically  summarized and 
interpreted in order to determine the 
corresponding  scenario, in terms of sliding states, as schematically depicted 
in Fig. \ref{fig:App_A}. (Note that here, for reasons of clarity, we show 
only a smaller subset of values of $\beta$.) 
%%%%%%%%%%%%%%%%%%%%%%%%%%%%%%%%
%%%%%%%%%%%%%%%%%%%%%%%%%%%%%%%

In brief: each value $\beta$ gets allocated an entry in the ceiling and floor 
rows. If the corresponding phase portrait reveals a sliding state, the 
entry is 
``checked'' (illustrated in the figure by ``x''), otherwise it is left 
unchecked. 
In the next step, if there are checked entries in the floor (ceiling) row, the 
corresponding row is marked for a sliding state as ``blue'' (floor) or 
``gold'' 
(ceiling). If only one of the colors is present, as in the panels 
(a) and (b) in Fig. \ref{fig:App_A}, or if no color is present, the 
procedure ends and the point $(F,T)$ is classified as belonging to the 
scenarios ``sliding states only at the ceiling (floor)'' or ``no sliding 
state'', respectively. If both colors are present, the color is set to 
``green'' 
(see panels (c) and (d) in Fig. \ref{fig:App_A}); if there is overlap between 
the checked entries in the top and in the bottom rows (panel (d)), the 
color is further 
changed to ``red'', otherwise it is kept unchanged. The state is then 
classified as belonging to the scenarios ``sliding states at both walls, but for 
non-overlapping  subsets of $\beta$'' (green) or ``coexisting sliding 
states'' (red), respectively.

\section{\label{sec:appB}Steady states of ``hovering''}

\begin{figure*}[!htbp]
\centering
\includegraphics[width=0.9\linewidth]{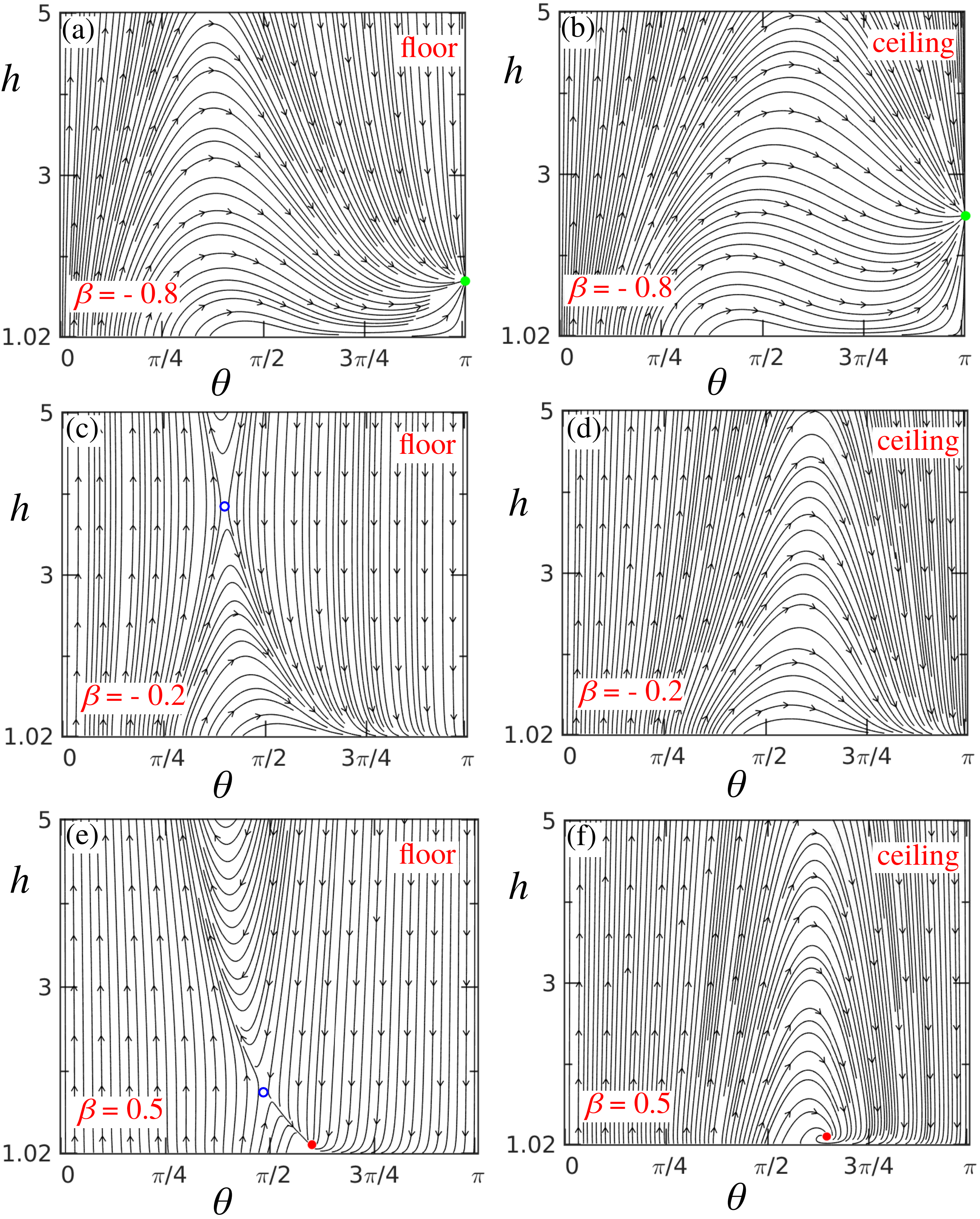}
\caption{{\small Phase portraits for a model heavy, gyrotactic 
Janus particle with constant flux chemical activity, $b_c < 0$, and 
the parameters $F=4,\;T=1$, near the floor (left column)] and near 
the ceiling 
(right column) for values 
$\beta=-0.8\;[(a),(b)],\;-0.2\;[(c),(d)]\;\text{and}\; 0.5\;[(e),(f)]$. The 
\textit{green} markers denote hovering states, while the \textit{red} 
markers denote sliding states; also the locations of saddle points 
are indicated by the open blue symbols in 
(c) and (e). Note that only the region $h 
\leq 5$ is shown, both for reasons of clarity and because for $h > 5$ the 
dynamics is basically that of a particle in the bulk.
}
}
\label{fig:App_B_1}
\end{figure*}
%%%%%%%%%%%%%%%%%%%%%%%%%%%%%%
As noted in the main text, it is known that the dynamics of a self-phoretic 
Janus particle near a wall exhibits an additional type of steady-state, the 
so called ``hovering'' state\cite{Uspal2015a}: the particle is at rest at 
a height $h^*$ above the wall and with an orientation $\theta^* = \pi$, 
while still 
pumping the fluid. Although there is only indirect experimental evidence for 
such states \cite{Uspal2018c}, it is nevertheless interesting to understand if 
these states, which for, e.g., heavy, gyrotactic particles, are favored 
at one wall (e.g., the ceiling), but disfavored at the other (the 
floor), can occur for 
the model particles studied here. 

%%%%%%%%%%%%%%%%%%%%%%%%%%
 \begin{figure}[!t]
\centering
\includegraphics[width=0.99\columnwidth]{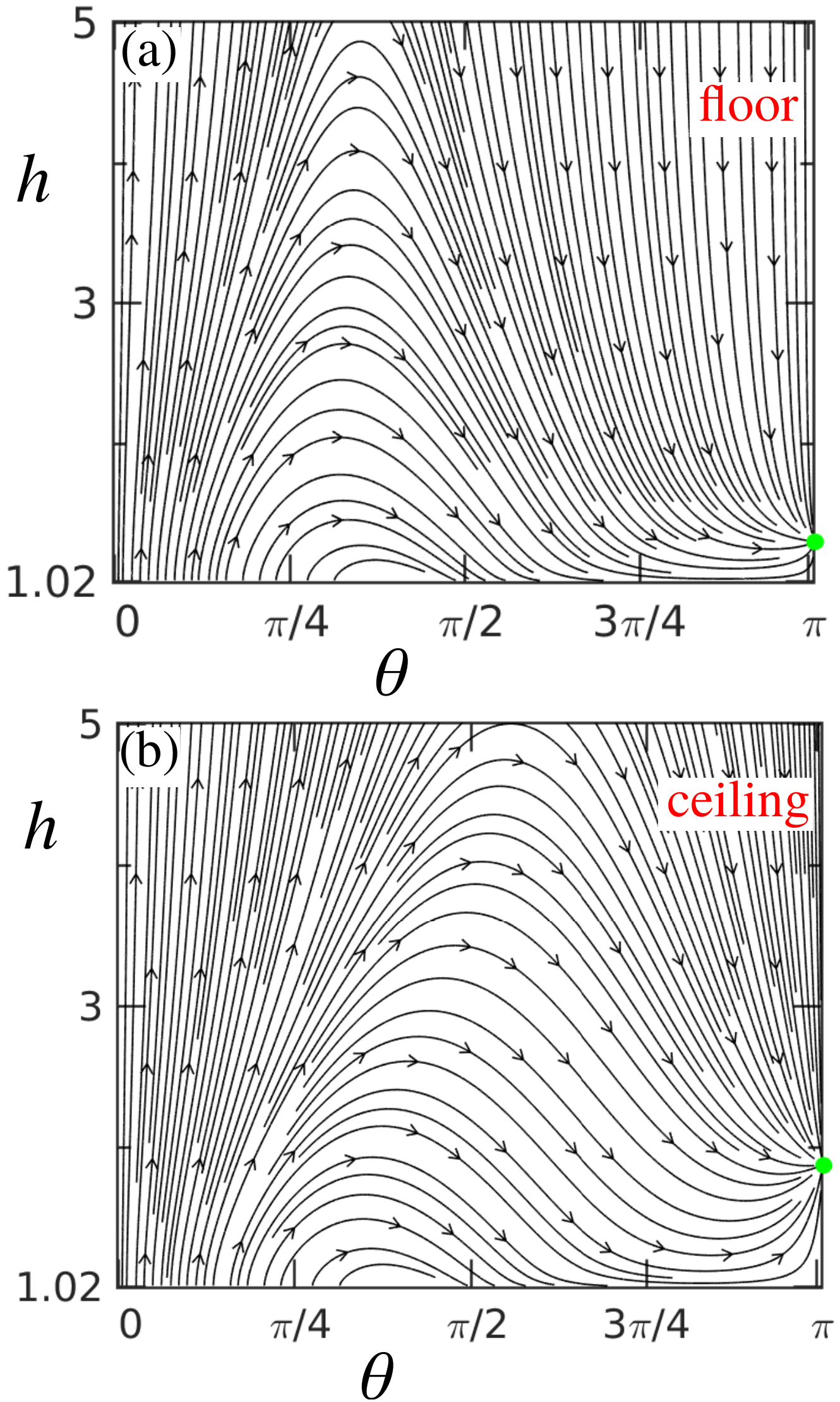}
\caption{{\small Phase portraits for a model heavy, gyrotactic Janus 
particle 
with variable flux chemical activity, $b_c < 0$, and the parameters 
$(F=4,\;T=1,\;\beta = -2.5)$ near the floor (a) and near the ceiling 
(b). In both cases, the 
\textit{green} markers denote hovering states. Note that only the 
region $h \leq 5$ is shown, both for reasons of clarity and because for $h > 5$ 
the dynamics is basically that of a particle in the bulk.
}
}
\label{fig:App_B_2}
\end{figure}
%%%%%%%%%%%%%%%%%%%%%%%%%%%
As shown in Fig. \ref{fig:App_B_1}, this is indeed the case here: similar to 
the 
case of the sliding states discussed in the main text, hovering states can also 
occur, both at the floor and at the ceiling, and coexistence of floor- and 
ceiling-hovering is possible. As illustrated in that figure, a model heavy, 
gyrotactic Janus particle with constant flux chemical activity, $b_c < 0$, and 
parameters $(F=4,\;T=1)$ exhibits sliding states for 
sufficiently large, positive 
values of $\beta$ (see Figs. \ref{fig:App_B_1} (e) and (f)). However, at 
sufficiently large, negative values of $\beta$ one observes hovering states 
(see Figs. \ref{fig:App_B_1} (a) and (b)). In both the floor and the 
ceiling case, the transition from hovering to 
sliding attractors is not continuous, at least within the limits of the 
numerical accuracy and of the constraint $h > 1.02$ imposed on the model. 
However, this transition involves a range of values for the parameter $\beta$ 
(see Figs. \ref{fig:App_B_1} (c) and (d)) within which steady motion of 
the particle is possible only in the bulk (far from the wall). 
Concerning the location of the hovering state, one notices that, in 
the case of coexistence of such states, the hovering occurs 
further away from the wall at the ceiling as compared to the floor. This 
follows from the fact that the particle is, due to the choice of the 
parameters, heavy 
(i.e., it tends to sediment near the floor). Accordingly, at 
the ceiling the gravity-induced velocity opposes the approach to the wall due 
to the phoretic velocity.

Hovering steady-states --- as well as the co-existence of such states --- 
occur also in the case of the variable flux chemical activity. 
This situation is illustrated by Fig. \ref{fig:App_B_2} which corresponds to 
the same parameters $(F,T)$ as above and to $b_c < 0$, but $\beta = 
-2.5$. As noted in the discussion in the main text, the two models are 
qualitatively equivalent concerning the type of steady-state scenarios.

%%%%%%%%%%%%%%%%%%%%%%%%%%
 \begin{figure}[!b]
\centering
\includegraphics[width = 
0.99\columnwidth]{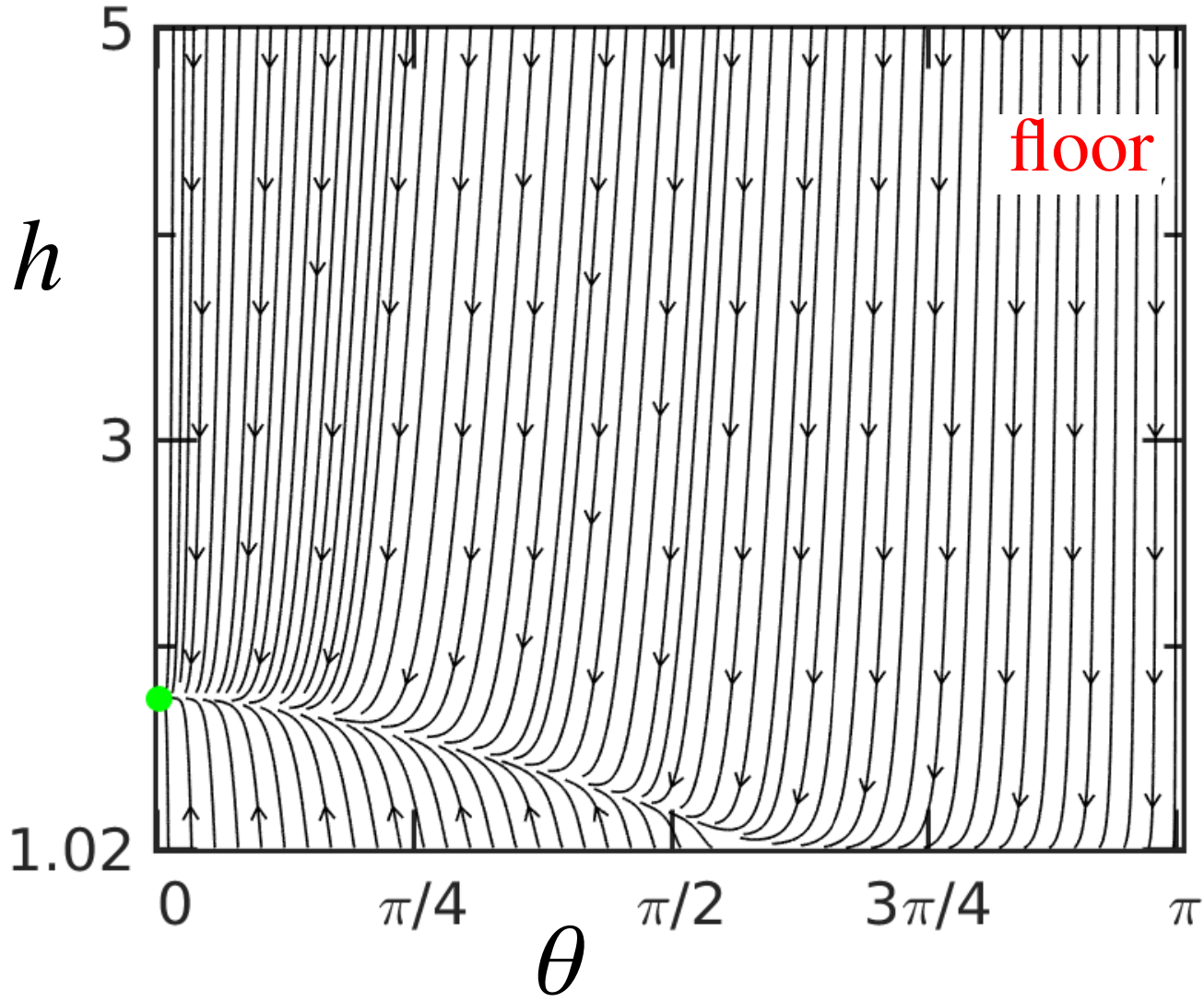}
\caption{
{\small 
Phase portrait for a model heavy, gyrotactic Janus particle with constant 
flux chemical activity, $b_c < 0$, and the parameters $(F=25,\;T=2,\;\beta = 
0.9)$ near the floor. The \textit{green} marker indicates the cap-down 
($\theta = 0$) hovering state. Note that only the region $h \leq 5$ is shown, 
both for reasons of clarity and because for $h > 5$ the dynamics is basically 
that of a particle in the bulk.
}
}
\label{fig:App_B_3}
\end{figure}
%%%%%%%%%%%%%%%%%%%%%%%%%%%
%%%%%%%%%%%%%%%%%%%%%%%%%%
 \begin{figure*}[!h]
\centering
\includegraphics[height = 
0.7\textheight]{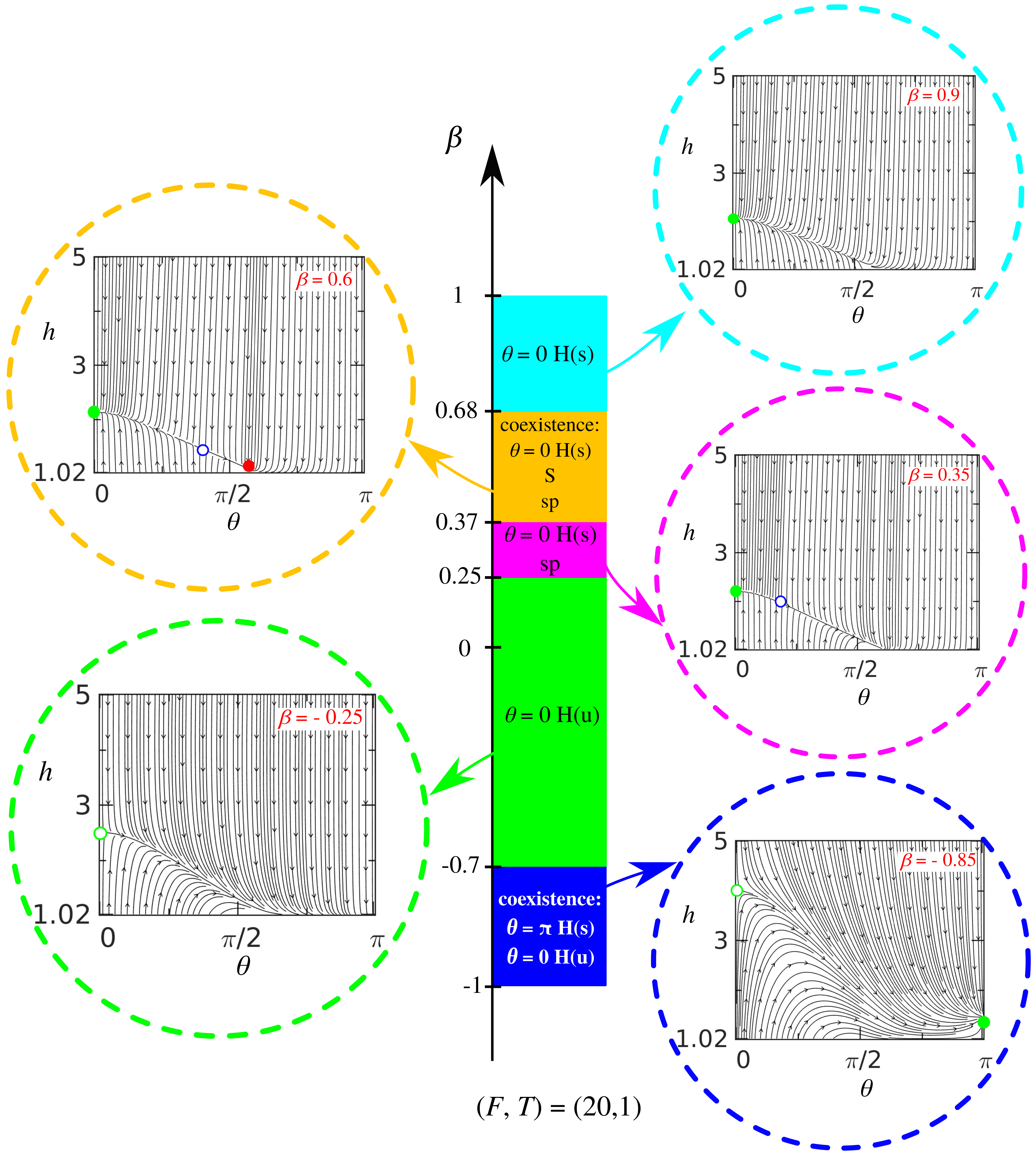}
\caption{
{\small 
Five possible phase portraits for a model heavy, gyrotactic Janus particle  
with constant flux chemical activity, $b_c < 0$, and the parameters $(F = 
20,\;T = 1)$, when moving near the floor. Upon varying the ratio $\beta$ of the 
phoretic mobilities, the following scenarios emerge (illustrated in the panels 
within the dashed circles). For $0.68 \lesssim \beta < 1$ (cyan region), a 
cap-down ($\theta = 0$, stable (s) hovering (H) state (filled green symbol) 
emerges. Decreasing $\beta$ further, in the range $0.37 \lesssim \beta 
\lesssim 0.68$ (orange region) one observes coexistence between the cap-down 
stable hovering state (filled green symbol) with a stable sliding (S) state 
(filled red symbol), separated by an (unstable, stationary) saddle point 
(sp, open blue symbol). The sliding state disappears once $\beta \simeq 0.37$, 
and for $0.25 \lesssim \beta \lesssim 0.37$ (magenta region) the cap-down 
stable hovering state (filled green symbol) coexists with the saddle point 
(open blue symbol), the location of which is shifting towards $\theta = 0$ as 
$\beta$ decreases. At $\beta \simeq 0.25$ the hovering point and the saddle 
point collide, and for $-0.7 \lesssim \beta \lesssim 0.27$ (green region) the 
phase portraits exhibit only an unstable (u) cap-down hovering state (open 
green symbol). Finally, at $\beta \simeq -0.7$ an 
additional, stable, cap-up ($\theta = \pi$) hovering state emerges (filled 
green symbol), and for $-1 \lesssim \beta \lesssim -0.7$ (blue region) 
coexistence of this state with the unstable cap-down state (open green symbol) 
is observed.
}
}
\label{fig:App_B_4}
\end{figure*}
%%%%%%%%%%%%%%%%%%%%%%%%%%% 
The hovering states discussed above are 
configurations with the active cap 
oriented \textit{away} from the wall (i.e., $\theta^* = \pi$), similar to those 
discussed in Ref. \citenum{Uspal2015a} in the absence of gravity. However, for 
the gyrotactic, bottom heavy, sedimenting active Janus spheres discussed here a 
new type of hovering state exists at the floor, with the active cap 
oriented 
\textit{towards} the wall (i.e., $\theta^* = 0$), as shown in Fig. 
\ref{fig:App_B_3}. This state is easy to grasp intuitively, in that it occurs 
if 
at a certain height $h^*$ above the wall the sedimentation velocity is exactly 
balanced by the self-propulsion (both of them depending on $h$ and, for 
$\theta = 0$, having opposite signs).

Finally, we note that for a sedimenting, gyrotactic, chemically active Janus 
particle the dynamic behavior at the floor is significantly enriched compared 
to the case, studied theoretically in Refs. \citenum{Uspal2015a} 
and \citenum{Bayati2019}, in 
which the effects of gravity (sedimentation and gyrotaxis) are disregarded. This 
enrichment occurs not only in terms of the appearance of a new type of 
steady state, i.e., the 
``cap down'' hovering noted above, but, as shown in Fig. \ref{fig:App_B_4}, 
also in 
terms of coexisting steady states of sliding and hovering at the floor 
corresponding to bi-stability. For example, for a Janus particle with 
parameters $(F,\;T) = (20,1)$ and 
$b_c < 0$, for $\beta < - 0.7$ one observes coexistence of a stable cap-up 
($\theta^* = \pi$) hovering state  with an unstable cap-down ($\theta^* = 0$) 
one. Upon increasing $\beta$ above the value $\beta \simeq -0.7$, the stable 
hovering point disappears; once $\beta > 0.25$, the unstable cap-down 
hovering 
state splits into a stable cap-down hovering one and a saddle-point. From $\beta 
> 0.37$, a sliding state also occurs, and for $0.37 \lesssim \beta \lesssim 
0.68$ one observes coexistence of a sliding state (cap slightly tilted away from 
the wall, $\theta^* > \pi/2$) with a stable cap-up hovering state ($\theta^* = 
0$), separated by a saddle point. Finally, for $\beta > 0.68$ only the stable 
cap-up hovering state is present in the dynamics.

An exhaustive investigation of this very rich dynamics is beyond the scope 
of this study and is left for future analysis.

\clearpage

\begin{onecolumn}

\section{\label{sec:appC}{Glossary and list of symbols}}

In order to facilitate the reading of the paper, here we summarize 
the mathematical symbols and abbreviations used throughout the main text of 
this study.

\begin{table*}[!h]
\label{tab:2}
\begin{tabular}{|c|l|}
\hline
\textbf{symbols} & \textbf{details} \\ \hline
\multicolumn{1}{|l|}{}          &                                                
                                                  \\
R                             & particle radius (m) (before Eq. 
(\ref{eq:Laplace}))                                                             
  
    \\
$U_0$                         & characteristic particle velocity (m/s) (Eq. 
(\ref{eq:Ufs}))                                         \\
$t_0$                         & characteristic time (s) (Eq. 
(\ref{eq:part_mot_gen}))                                                        
  
   \\
$F_0$                         & characteristic force (N) (Eq. 
(\ref{eq:scaled_Fg}))                                            
            \\
$T_0$                         & characteristic torque (N $\times$ m)  (Eq. 
(\ref{eq:scaled_Tg}))                                            
         \\
$\mathbf{U}$                  & particle translational velocity (m/s) 
(Eq. (\ref{eq:BC_u_part}))                                          \\
$t$                           & time elapsed (s) (Eq. \ref{eq:part_mot_gen}) 
                                                              \\
${\mathbf F}_{\textit g}$     & gravitational force (N) (Eq. 
(\ref{eq:scaled_Fg}))                                            
             \\
${\mathbf T}_{\textit g}$     & gravitational torque (N $\times$ m) 
(Eq. 
(\ref{eq:scaled_Tg}))                                            
           \\
$h$                           & dimensionless height of the particle 
centroid from the wall (Fig. \ref{fig3})        \\
$D$                           & diffusion coefficient of the solute molecules 
(m$^2$/s) (Eq. (\ref{eq:BC_c_part}))                          \\
$Pe$                          & dimensionless P\'eclet number (page 6, before 
Eq. (\ref{eq:Laplace}))                         \\
$\mathbf{r}$                  & position vector of any point in the solution 
(m) (Eq. (\ref{eq:Laplace}))                                  \\
${\mathbf r}_s$               & position vector at a point on the particle 
surface (m) (Eq. (\ref{eq:BC_c_part}))                           \\
$c(\mathbf{r})$               & solute number density at a point in the 
solution (m$^{-3}$) (Eq. (\ref{eq:Laplace}))                      \\
$\mathcal{Q}$                 & total rate of solute production 
(m$^{-2} \times$ s$^{-1}$) (Eq. (\ref{eq:Laplace}))                            
\\
$\mathcal{S}$                 & particle surface (Fig. \ref{fig1})               
                                                         \\
$\mathbf{p}$                  & unit vector along the particle symmetry 
axis, towards the active pole (Fig. \ref{fig3})   \\
$q({\mathbf r}_s)$            & chemical activity function (Eq. 
(\ref{eq:choices_f}))                                                          
\\
$\mathbf{e}_{x,y,z}$                  & unit vector along $x,y,z$ direction 
(Fig. 
\ref{fig3}) \\
$\mathbf{n}$                  & unit vector denoting the outer normal of 
the wall (Fig. \ref{fig3})                                     \\
${\mathbf v}_s$               & phoretic slip velocity (m/s) (Eq. 
(\ref{eq:phor_slip}))                                                    \\
$b({\mathbf r}_s)$            & surface mobility of the particle (m$^5$/s) 
(Eq. (\ref{eq:phor_slip}))                                        \\
Re                            & dimensionless Reynolds number (between Eqs. 
(\ref{eq:def_b}) and (\ref{eq:Stokes}))                                         
\\
${\mathbf v}({\mathbf r})$    & velocity field (m/s) (Eq. (\ref{eq:Stokes}))   
                                                         \\
$P(\mathbf{r})$               & fluid pressure field (Pa) (after Eq. 
(\ref{eq:Stokes}))                                                \\
$\mathbb{M}$                  & mobility matrix (Eq. 
(\ref{eq:mob_matrix_connect}))                                                   
                 \\
$m_{i,j}$                     & elements of the mobility matrix (Eq. 
(\ref{eq:scaled_Ms}))                                          
        \\
$\mathcal{U}(\beta)$          & a prefactor depending on the 
activity-based-model (Eq. (\ref{eq:Ufs}))                      \\
$\mathbf{u}$                  & dimensionless particle translational velocity 
(Eq. (\ref{eq:scaled_us}))      \\
$F$                           & dimensionless apparent weight of the particle 
(Eq. (\ref{param_F}))                                     \\
$T$                           & dimensionless gravitactic torque on the particle 
(Eq. (\ref{param_T}))                                  \\
$\mathcal{F}$                 & $F/9$ (Eq. 
(\ref{eq:scaled_Fg})) \\
$\mathcal{T}$                 & $T/12$ (Eq. 
(\ref{eq:scaled_Tg})) \\
$m_a$                         & mass of the deposited catalyst layer (kg) (Eq. 
(\ref{eq:def_delta}))                                          \\
g                             & acceleration due to gravity (m$^2$/s) (Fig. 
\ref{fig1})                                                 \\
$\hat{\mathcal{I}}$           & identity tensor (after Eq. (\ref{eq:phor_slip}))
\\ \hline
\end{tabular}
\end{table*}

\begin{table*}[!h]
\label{tab:3}
\begin{tabular}{|c|l|}
\hline
\textbf{Greek}        & \textbf{details}                                    
                      \\ \hline
\multicolumn{1}{|l|}{}          &                                                
                                                  \\
$\mu$                         & dynamic viscosity of the suspending fluid (Pa 
$\times$ sec) (before Eq. (\ref{eq:Stokes}))                    \\
$\Omega_0$                    & characteristic particle rotational velocity 
(s$^{-1}$) (before Eq. (\ref{eq:def_beta}))               \\
$\mathbf{\Omega}$             & particle rotational velocity 
(s$^{-1}$) (Eq. (\ref{eq:BC_u_part}))
\\
$\theta$                      & orientation angle of the particle (Fig. 
\ref{fig3})                                                       \\
$\delta$                      & maximum thickness of the catalyst layer (m) 
(Fig. \ref{fig4})                                            \\
$\chi$                        & angle between ${\mathbf r}_s$ and director 
$\mathbf{p}$ (Fig. \ref{fig3})                                 \\
$\nabla$                      & gradient operator (Eq. 
(\ref{eq:BC_c_part})) \\
$\nabla_{||}$                 & surface gradient operator (Eq. 
(\ref{eq:phor_slip}))                                                           
\\
$\boldsymbol{\hat{\sigma}}$       & stress tensor (N/m$^2$) (Eq. 
(\ref{eq:Stokes}))                                                             
\\
$\beta$                       & ratio of inert and active phoretic mobilities 
(Eq. (\ref{eq:def_beta}))                                      \\
$\boldsymbol{\omega}$             & dimensionless particle rotational velocity 
(Eq. (\ref{eq:scaled_oms}))    \\
$\tau$                        & dimensionless time (Eq. 
(\ref{eq:dimless_motion_part}))                                  \\
$\rho_a$                      & mass density of the catalyst (kg/m$^3$) 
(Eq. 
(\ref{param_F}))                                           \\
$\rho_s$                      & mass density of the inert core of the particle 
(kg/m$^3$) (Eq. (\ref{param_F}))                         \\
$\rho$                        & mass density of the hydrogen peroxide solution 
(kg/m$^3$) (Eq. (\ref{param_F}))    
                     \\ \hline
\textbf{subscripts}   & \textbf{details}        \\ \hline
\multicolumn{1}{|l|}{}  &                                                \\
$rep$                 & reported value                                 \\
$cf$                  & constant flux                                  \\
$vf$                  & variable flux                                  \\
$c$                   & active cap                                     \\
$i$                   & inert portion of the particle surface               
\\
$x,y,z$               & x,y,z Cartesian coordinates           \\
$w$                   & wall (floor or ceiling)                        \\
$g$                   & gravitational contribution                     \\
$ph$                  & phoretic contribution                          \\
$O$                   & position of the hcenter of mass of the particle \\ 
\hline
\textbf{superscripts} & \textbf{details}        \\ \hline
\multicolumn{1}{|l|}{}  &                                                \\
(1,0)                 & $(b_c,b_i)=(1,0)$                              \\
(0,1)                 & $(b_c,b_i)=(0,1)$                              \\
($fs$)                & free space                                     \\ \hline
\end{tabular}
\end{table*}

\clearpage

\section{\label{sec:appD}Description of the video files of Pt/PS and 
Pt/SiO$_2$ active Janus particles}

Videos of Pt/PS and Pt/SiO$_2$ active Janus particles:
\begin{itemize}
 \item The video files SI.V1 to SI.V5 illustrate the motion of platinum-capped 
(Pt 
thickness $\delta =  8.5$ nm) polystyrene (Pt/PS) Janus particles of diameters 
2.4 and 5 $\mu$m, respectively, in 3\% (v/v) aqueous H$_2$O$_2$ solution both 
at the ceiling and at the floor.
  \item The video files SI.V6 and SI.V7 illustrate the motion of 
platinum-capped (Pt 
thickness $\delta =  8.5$ nm) silica (Pt/SiO$_2$) Janus particles of diameters 
1 and 4 $\mu$m, respectively, in 3\% (v/v) aqueous H$_2$O$_2$ solution
at the floor. The video recording was carried out with a u-eye 2240c camera at 
a 
rate of 10 frames per second (fps). The yellow curves show tracked trajectories 
of particles (tracking has been performed using the ImageJ software and an 
in-house Matlab script).
\end{itemize}

\noindent \textbf{Video-file details:}\newline

\noindent VideoSI.V1: Motion of three Pt/PS Janus particles of 2.4 $\mu$m 
diameter, at the floor (file name: 
\verb!VideoSI.V1_2.4um_ptps_3H2O2_bottom_20x.m4v!).\newline

\noindent VideoSI.V2: Motion of four Pt/PS Janus particles of 2.4 $\mu$m 
diameter, at the ceiling (file name: 
\verb!VideoSI.V2_2.4um_ptps_3H2O2_top_20x.webm!).\newline

\noindent VideoSI.V3: Motion of two Pt/PS Janus particles of 5 $\mu$m diameter, 
at the floor (file name: 
\verb!VideoSI.V3_5um_ptps_3H2O2_bottom_20x.webm!).\newline

\noindent VideoSI.V4: Motion of three Pt/PS Janus particles of 5 $\mu$m 
diameter, at the ceiling. The third particle arrives at the ceiling at time t = 
14 s (file name: \verb!VideoSI.V4_5um_ptps_3H2O2_top_20x.webm!).\newline

\noindent VideoSI.V5: Motion of two Pt/PS Janus particles of 2.4 $\mu$m 
diameter, at the ceiling. The third particle arrives at the ceiling at time t = 
30 s (file name: \verb!VideoSI.V5_2.4um_ptps_3H2O2_top_20x.webm!).\newline

\noindent VideoSI.V6: Motion of three Pt/SiO$_2$ Janus particles of diameter 
1 $\mu$m, at the floor 
(file name: \verb!VideoSI.V6_1um_ptSiO2_3H2O2_bottom_20x.m4v!).\newline

\noindent VideoSI.V7: Motion of six Pt/SiO$_2$ Janus particles of diameter 
1 $\mu$m, at the floor (file name: 
\verb!VideoSI.V7_4um_ptSiO2_3H2O2_bottom_20x.webm!).

\end{onecolumn}

\end{document}